\documentclass[amsmath,amssymb,prl,aps,superscriptaddress, floatfix,tightenlines,showpacs,twocolumn]{revtex4-2}
\usepackage{graphicx}
\usepackage{dcolumn}
\usepackage{bm}
\usepackage{dutchcal}
\usepackage{color}

\usepackage{booktabs}
\usepackage{float}
\usepackage[hyperindex,breaklinks]{hyperref}

\def\Im {\mbox{Im}}

\def\be{\begin{equation}}       \def\ee{\end{equation}}
\def\bea{\begin{eqnarray}}      \def\eea{\end{eqnarray}}
\def\bp{\begin{pmatrix}} \def\ep{\end{pmatrix}}
\def\beaa{\begin{equation}\begin{aligned}}
		\def\eeaa{\end{aligned}\end{equation}}

\def\nn{\nonumber}

\usepackage{verbatim}

\begin{document}
\title{An Explicit Wavefunction of the Interacting Non-Hermitian Spin-1/2 1D System}
\author{Yue Wang}
\affiliation{Department of Physics, Zhejiang University, Hangzhou 310027, China}
\affiliation{New Cornerstone Science Laboratory, Department of Physics, School of Science, Westlake University, Hangzhou 310024, Zhejiang, China}
\author{Xiangyu Zhang}
\affiliation{New Cornerstone Science Laboratory, Department of Physics, School of Science, Westlake University, Hangzhou 310024, Zhejiang, China}
\author{Zhesen Yang}
\email{yangzs@xmu.edu.cn}
\affiliation{Department of Physics, Xiamen University, Xiamen 361005, Fujian Province, China}
\author{Congjun Wu}
\email{wucongjun@westlake.edu.cn}
\affiliation{New Cornerstone Science Laboratory, Department of Physics, School of Science, Westlake University, Hangzhou 310024, Zhejiang, China}
\affiliation{Institute for Theoretical Sciences, Westlake University, Hangzhou 310024, Zhejiang, China}
\affiliation{Key Laboratory for Quantum Materials of Zhejiang Province, School of Science, Westlake University, Hangzhou 310024, Zhejiang, China}
\affiliation{Institute of Natural Sciences, Westlake Institute for Advanced Study, Hangzhou 310024, Zhejiang, China}


\begin{abstract}
We present an explicit Bethe-ansatz wavefunction to a 1D spin-$\frac{1}{2}$ interacting fermion system, manifesting a many-body resonance resulting from the interplay between interaction and non-Hermitian spin-orbit coupling. 
In the dilute limit, the Bethe-ansatz wavefunction is factorized into Slater determinants and a Jastrow factor.
An effective thermodynamic distribution is constructed with an effective Hamiltonian including a repulsion resulting from Pauli's exclusion principle and a distinctive zigzag potential arising from the resonance. 
The competition between these effects leads to a transition from a uniformly distributed configuration to a phase separation.
Clustering of particles with identical spins is observed in the latter phase, demonstrating that the many-body resonance effect is enhanced by the repulsive interaction.

\end{abstract}
\maketitle

{\em Introduction.}---
The non-Hermitian skin effect (NHSE) has attracted significant attentions recently \cite{lee2016anomalous, martinez2018non, yao2018edge, lee2019hybrid, lee2019anatomy, borgnia2020non, SuPeng2020Edge, SuPeng2024ChiralSkin}. 
The associated exotic properties, such as the complex-valued spectrum and the localization on boundaries, can be described by the theory of the generalized Brillouin zone, in which momenta are complex-valued \cite{yao2018edge,Emil2018BulkBoundary, Murakami2019Band, Yang2020BulkBoundary, Yao2018ChernBand, Song2019OpenSys, Yang2020Correspondence}.
These distinctive features are highly sensitive to boundary conditions.
For example, the eigenstates in non-Hermitian systems are extended under the periodical boundary condition (PBC) while localized under the open boundary condition (OBC), which contrasts to the case in Hermitian physics. 
Experimentally, the NHSE has been observed in various systems, including metamaterials \cite{brandenbourger2019non, ghatak2020observation}, photonic systems \cite{xiao2020non}, electrical circuits \cite{hofmann2020reciprocal, Zhang2021Circuit}, acoustic crystals \cite{zhang2021acoustic}, and cold atomic systems \cite{Yanbo2022ColdAtom}.

Despite the significant progress in the NHSE, current studies  predominantly focus on the single-body physics.
How the NHSE behaves under strong interactions remains an open question, and non-perturbative analytical studies are desired. 
The Bethe-ansatz (BA) method \cite{korepin1997quantum, wang2015off} is a systematic tool for studying one dimensional (1D) integrable systems, including the Lieb-Liniger model of the interacting Bose gas \cite{Lieb1963BoseGasI, Lieb1963BoseGasII}, the Gaudin-Yang model of the interacting Fermi gas \cite{Yang1967FermiGas, GAUDIN1967FermiGas}, and the Lieb-Wu solution to the Hubbard model \cite{Lieb1968Hubbard}.
When applied to non-Hermitian systems \cite{SciPostPhys.8.3.044, New.J.Phys.22_123040, PhysRevResearch.6.L032067}, it has been found that NHSE is suppressed by repulsive interactions \cite{Masahito2021DissipatHubbard, Pan2023BoseGas, WZhong2023SpinChain, Kattel2023SpinChain, CShu2024BoseHubbard}.
However, the complexity of BA wavefunctions makes it difficult to calculate observables.
It would be desired to construct an explicit  many-body wavefunction to facilitate a deeper understanding of the NHSE in interacting systems, akin to the Ogata-Shiba-type and the Laughlin-type wavefunctions \cite{ogata1990, laughlin1983anomalous, girardeau1960relationship}.

In this work, we present a concise expression for the many-body wavefunction in a 1D spin-1/2 fermion system with the non-Hermitian spin-orbit coupling (SOC).
As a result of the repulsive $\delta$-interaction, each particle behaves as a soft boundary to particles with opposite spins, inducing an effective unidirectional attraction between them. 
Resonance states are formed here instead of bound states, {\it i.e.}, the Bethe string states \cite{Wu2019String, Wu2018ExString, Takahashi}.
The explicit many-body wavefunction is constructed in the dilute limit, which is a rare example in many-body physics.
It consists of the product of Slater determinants and the Jastrow factor \cite{Jastrow} reflecting the resonance between particles with opposite spins. 
Remarkably, a phase transition occurs as the interacting strength increases, where the level of localization exhibits a jump, indicating that the resonance is enhanced by the repulsive interaction.

{\em Model.}---We start with the following 1D non-Hermitian many-body Hamiltonian with system length $L$ ($\hbar=1$)
\begin{equation}
    \hat{H} = \sum_{l=1}^N \frac{\left(-i\nabla_l+i m \alpha \sigma_l^z\right)^2}{2 m} + 
    \sum_{ l < l'} 2g \delta\left(x_l-x_{l'}\right),
\end{equation}
where $\alpha>0$ and $g>0$ represent the strength of the non-Hermitian SOC and the repulsive interaction respectively; 
the PBC is assumed. 
Since the $z$-component of total spin is conserved, the eigenvalues and eigenstates can be labelled by the particle numbers of two components, {\it i.e.}, $N_\uparrow$ and $N_{\downarrow}$. 
In the following, the real and imaginary parts of the complex momentum are defined as
\begin{eqnarray}
k_{l,\sigma_l}=(\chi_{l,\sigma_l}
+i \mathcal{\eta}_{l,\sigma_l}
)/L,
\end{eqnarray}
where $\sigma_l=\pm 1$ represents the spin $z$-component.

We warm up by considering the single-body problem. 
If the particle carries spin $\sigma$ ($\sigma=\uparrow,\downarrow$), the eigenstates are denoted as $e^{i k_{\sigma}x}|\sigma\rangle$, and the eigen-energies are $(k_{\uparrow,\downarrow}\pm im \alpha)^2/(2m)$. 
The momentum is quantized as $k_\sigma=2 \pi n_\sigma/L$ under the PBC. 
Upon the OBC, the spin-up and down particles localize at the right and left boundaries, respectively.
The single-particle localization length is $\lambda_s=(m\alpha)^{-1}$, which is independent of the system size, indicating the presence of bound states.
This is the conventional NHSE discussed in the literature.



{\em Two-body case.}---
With $N_\uparrow=2$ and $N_\downarrow=0$, the corresponding eigenstate is a Slater determinant of plane wave states with $k_{i, \uparrow}=2\pi n_{i, \uparrow}/L$ and $i=1,2$. 
The $\delta$-interaction does not manifest here due to Pauli's exclusion principle.
The corresponding eigenenergy is $E=\sum_i(k_{i,\uparrow} +im \alpha)^2/2m$. 
The case with two spin-down particles can be constructed in parallel. 

Non-trivial interaction effect emerges with a pair of particles of opposite spins.
The eigenstate is written as $\varphi(x_1, x_2) | \! \! \downarrow \uparrow  \rangle - \varphi(x_2, x_1) | \! \! \uparrow  \downarrow\rangle$.
Using the center of mass coordinate $X=(x_1+x_2)/2$ and the relative coordinate $r=x_1-x_2$, the wavefunction is decomposed as  $\varphi(x_1,x_2)=\Phi(X) \phi(r)$, satisfying the following equations,
\begin{equation}
\begin{array}{l}
\left(-\frac{\nabla_X^2}{4m} -m \alpha^2\right) \Phi(X)=E_X \Phi(X), \vspace{1ex} \\
\left(-\frac{\nabla_r^2}{m} - 2\alpha \nabla_r + 2g \delta(r)\right) \phi(r)=E_r \phi(r) .
\end{array}.
\label{E3}
\end{equation}
$\Phi(X)$ is solved as $e^{i K_X X}$ where $K_X \! = \! 2\pi n_{K}/L$.
The non-Hermitian term $-2\alpha \nabla_r$ only appears in the motion of the relative coordinate, where the $\delta$-potential acts as a soft boundary.
As a result, the reminiscence of the NHSE would bring an effective unidirectional attraction between two particles, explained as follows. 
The relative motion is solved as
\begin{equation}
    \phi(r)=Ae^{ik_r r}+Be^{-i k_r r -2m\alpha r},
\label{eq:WV}
\end{equation}
where $A, B$ are scattering amplitudes, 
$k_r=(\chi_r+i \eta_r)/L$ is the complex momentum.
Matching wavefunctions on both sides of the $\delta$-potential, it yields,
\begin{align}
&\frac{\beta}{g}\Big( (-1)^{n_K} \operatorname{cosh} m \alpha L - \operatorname{cosh}m \beta L \Big) 
=\operatorname{sinh}m \beta L,
\label{equation to kr}
\end{align}
where $\beta = \alpha+ i\frac{k_r}{m}$.
As shown in Supplemental Material (SM) I, in the case of $L \gg \lambda_s$, 
Eq. (\ref{equation to kr}) is solved as
\begin{eqnarray}
\! \! \! \!
\chi_r=(2n_r + n_K)\pi, \
\mathcal{\eta}_r= \ln(1+\frac{g}{\alpha}), \ 
\frac{A}{B}=-(1+\frac{\alpha}{g}).
\label{E7}
\end{eqnarray}

$\phi(r)$ is identical to a single-body wavefunction of spinless particle subjected to a `soft' boundary condition, which lies between the cases of OBC and PBC, since the $\delta$-potential permits partial transmission.
In all cases, the solution possesses a pair of momenta, whose imaginary parts are summed to $2m\alpha$.
In the OBC case, both imaginary parts equal $m\alpha$, while in the PBC case, one becomes real and the other carries the imaginary part of $2m\alpha$.
In our case, a small imaginary part $\eta_r/L$ is at the order of $1/L$, and the other remains at the order of $2m\alpha$.
Consequently, the decay of $\phi(r)$ is at the length scale of $L$, such that these states are resonance rather than bound states \cite{li2020critical, li2021impurity}.
The situation of OBC is recovered for $g \sim \frac{1}{mL}e^{L/\lambda_s}$ in which case the localization length $L/\eta_r \sim \lambda_s$, while that of PBC corresponds to $g=0$.



In the lab frame, $\varphi(x_1,x_2)$ is written as,
\begin{equation}
\small
\begin{aligned}
    & \theta(x_1 > x_2)\left(A e^{i(k_{1,\downarrow}x_1 + k_{2,\uparrow}x_2)}
    + B e^{i(k_{2,\downarrow}x_1 + k_{1,\uparrow}x_2)}\right) \\
    &+ \theta(x_2 > x_1)\left(A'e^{i(k_{1,\downarrow}x_1 + k_{2,\uparrow}x_2)}
    + B' e^{i(k_{2,\downarrow}x_1 + k_{1,\uparrow}x_2)}\right)
    \label{Two-body wave function - no approx}
\end{aligned}
\end{equation}
where $0<x_{1,2}<L$.
$B$ and $B^\prime$-terms are the reflected waves of $A$ and $A^\prime$-terms respectively.
After the reflection, the real parts of momenta switch, but their imaginary parts change due to the SOC. 
As a result, the imaginary parts of $k_{1,\downarrow}$ and $k_{2,\uparrow}$ are at the order of $1/L$,
\begin{equation}
\begin{aligned}
k_{1,\downarrow}=
(\chi_{1}+i\mathcal{\eta}_r)/L,
\ \ \
k_{2,\uparrow}=
(\chi_{2}-i\mathcal{\eta}_r)/L,
\label{E9}
\end{aligned}\end{equation}
while that of the reflected momenta $k_{1,\uparrow}$ and $k_{2,\downarrow}$ become finite as
\begin{equation}
k_{1,\uparrow}= k_{1,\downarrow} -2im\alpha, 
\ \ \
k_{2,\downarrow}=k_{2,\uparrow} +2im\alpha.
\label{Diffractive reflection}
\end{equation}
Here $\chi_i=2\pi n_i$ with $i=1,2$.
$A^\prime$ and $B^\prime$-terms are the transmitted waves of $A$ and $B$-terms respectively.
The PBC yields $A'/A=e^{ik_{1,\downarrow}L}$ and $B'/B=e^{ik_{2,\downarrow}L}$.

We view $A$, $A^\prime$-terms as the incident waves and $B$, $B^\prime$-terms as the reflected waves.
Since the imaginary parts of their momenta behave differently, in the case of $L \gg \lambda_s$, the reflected waves can be dropped if the inter-particle distance exceeds $\lambda_s$.
Then the wavefunction is simplified to the product of plane-waves and a Jastrow factor,
\begin{equation}
    \varphi(x_1,x_2) = 
    A e^{i\left(\chi_{1} x_1 + \chi_{2} x_2\right)/L} \cdot 
    e^{-\frac{1}{2} W(x_1-x_2)},
    \label{pair wavefunction}
\end{equation}
where $e^{-\frac{1}{2} W(x_1-x_2)}$ is given by the sum of step-functions modified by the imaginary parts of the complex momenta
\begin{equation*}
\begin{aligned}
    e^{-\eta_r(x_1 - x_2)/L} 
    \left( 
    \theta (x_1 > x_2) +
    \theta (x_2 > x_1) 
    \left(1 + \frac{g}{\alpha}\right)^{-1}
    \right).
\end{aligned}
\end{equation*}
More explicitly, 
\begin{equation}
    W(r)= 2\eta_r \left( \frac{r}{L} + \theta(-r) \right), \ \ -L<r <L.
\end{equation}

\begin{figure}[tp]
\centering
\includegraphics[scale=0.19]{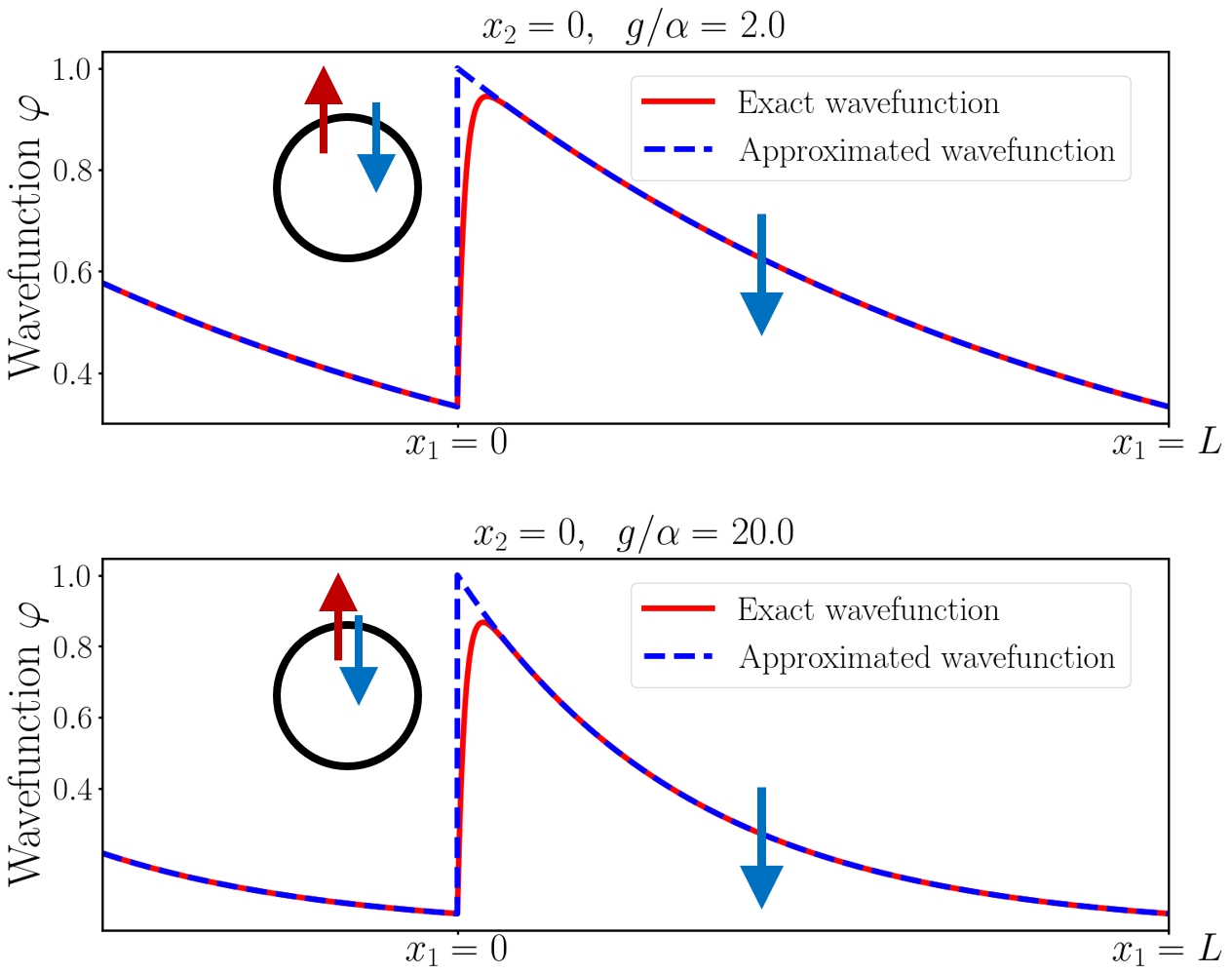}
\caption{The 2-body wavefunction $\varphi(x_1, x_2)$ with particles of opposite spins by fixing the spin-up one at $x_2=0$.
The parameter values are $\chi_{1}=\chi_{2}=0$, $m=1$, $\alpha=1$ and $L=50$.
The peak appears at $x_1 \sim \ln (L \lambda_s^{-1}) / ( L \lambda_s^{-1} )$.
In the case of $L \gg \lambda_s$, the peak is located at $x_1=x_2$ and $\varphi$ becomes discontinuous. 
It implies that the spin-down particle tends to lie on the right side of the spin-up particle, forming a resonant pair on the ring.
The wavefunction becomes more localized as the strength of repulsive interaction $g$ increases, which indicates that the resonance is enhanced by the repulsive interaction.
}
\label{FIG1}
\end{figure}

The exact and approximated wavefunctions Eq. (\ref{Two-body wave function - no approx}) and Eq. (\ref{pair wavefunction}) are shown in Fig. \ref{FIG1} respectively.
The spin-up particle is fixed at $x_2 \! = \! 0$.
Increasing $x_1$ from $0$, $\varphi(x_1,x_2)$ rapidly reaches the peak located at $x_{1,\rm peak} \sim \ln (L \lambda_s^{-1}) / (L \lambda_s^{-1})$.
If $L\gg \lambda_s$, $x_{1,\rm peak}$ coincides with $x_2$.
The peak is followed by a slow decay at the length scale of $L$, which means the spin-down particle prefers the right side of the spin-up one.
This can be understood as a weaker version of the NHSE with localization length $L/\eta_r$. 
Note that this length decreases as $g$ increases, indicating that the localization is enhanced by the repulsive interaction, which distinguishes it from the conventional NHSE suppressed by repulsions as reported in the previous studies \cite{Masahito2021DissipatHubbard, Pan2023BoseGas, WZhong2023SpinChain, Kattel2023SpinChain, CShu2024BoseHubbard}.
The behavior at $x_1<0$ can be obtained by applying the PBC. 
As we will explain later, the two-body wavefunction can be generalized to the many-body case where the above picture still holds. 


{\em Many-body problem: BA equations.}--- 
For eigenstates with $N_{\downarrow}$ down spins and $N_{\uparrow}=N-N_{\downarrow}$ up spins, the corresponding BA equations are:
\begin{equation}
    \begin{aligned}
    &\prod_{i=1}^{N_\downarrow} \!
    \frac{\Lambda_{i}-\overline{k}_{l}-i m g}
    {\Lambda_{i}-\overline{k}_{l}+i m g} \! = \!
    e^{i \overline{k}_{l} L} e^{m \alpha L}, \\
    &\prod_{l=1}^{N} \!
    \frac{\Lambda_{i}-\overline{k}_{l}-i m g}
    {\Lambda_{i}-\overline{k}_{l}+i m g} \! = \! - 
    e^{2 m \alpha L} \prod_{i'=1}^{N_\downarrow} \!
    \frac{\Lambda_{i}-\Lambda_{i'}-2 i m g}
    {\Lambda_{i}-\Lambda_{i'}+2 i m g},
    \label{BA equation}
    \end{aligned}
\end{equation}
where $\{\overline{k}_{l}\}$ and $\{\Lambda_{i}\}$ are $N$ and $N_\downarrow$ variables to be determined, with $1\leq l \leq N$ and $1 \leq i, i^{\prime} \leq N_{\downarrow}$.
Once $\{\bar{k}_{l}\}$ are obtained, the corresponding momentum of the $l$-th particle 
with spin $\sigma_l$ is given by
\begin{equation}
k_{l,\sigma_l} = \overline{k}_{l} - im\alpha\sigma_l.
\end{equation}
Note that the scattering is always non-diffractive in terms of $\overline{k}_l$, which ensures the integrability of the model.

\begin{figure}[tp]
  \centering
  \includegraphics[scale=0.2]{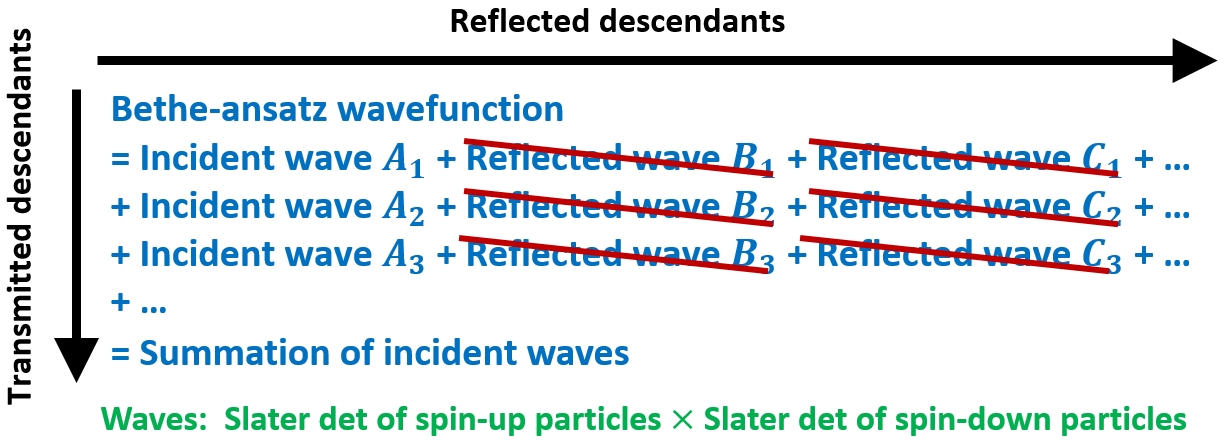}
  \caption{The construction of the Bethe-ansatz wavefunction.
  It is composed of many `plane waves'.
  Each plane wave is a product of Slater determinants, due to the Fermi statistics.
  The `plane waves' in the same column or the same row are connected by transmission or reflection respectively.
  Due to the diffractive reflection described by Eq. (\ref{Diffractive reflection}), the reflected descendants of $A_i$ are all suppressed.}
  \label{Illustration of BA wavefunction}
\end{figure}

The solutions to BA equations are significantly simplified due to the non-Hermitian SOC.
For example, in the two-body case, the solutions can be effectively expressed by the Eq. (\ref{E9}), since those reflected waves with momenta $\{k_{1,\uparrow}, k_{2,\downarrow}\}$ are removed in the approximation to the wavefunctions.
The same thing happens in the many-body case.
As illustrated in Fig. (\ref{Illustration of BA wavefunction}), the many-body wavefunction is composed of many `plane waves', connected with each other by scattering.
Those `plane waves' in the same column are connected by transmission, which share the same momenta since transmission exchanges the positions of the particles while preserving their momenta.
For example, $A_2$, $A_3\cdots$ are the transmission descendants of $A_1$, with momenta
\begin{eqnarray}
    k_{i,\downarrow}&=&
    (\chi_{i} + i\mathcal{\eta}_{r} N_\uparrow)/L, ~~ 
    k_{j,\uparrow}=(\chi_{j} - i\mathcal{\eta}_{r} N_\downarrow)/L.
    \label{Dominant channel momentum distribution}
\end{eqnarray}
Here $1 \leq i\leq N_\downarrow$, $N_\downarrow + 1 \leq j\leq N$, $\chi_l = 2 \pi n_l$ and $\mathcal{\eta}_r$ is defined in Eq.~(\ref{E7}).
$A_1$, $A_2$, $A_3\cdots$ are denoted as incident waves, like the $A$ and $A'$-terms in the two-body case.
Each incident wave has many reflected descendants, aligned in the same row. 
Due to the diffractive reflection described by Eq. (\ref{Diffractive reflection}), these descendants decay much faster than the incident waves, therefore can be discarded in the dilute limit defined as,
\begin{equation}
\frac{d}{\lambda_s }
\gg \ln \left(1+\frac{g}{\alpha}\right),
\label{dilute condition}
\end{equation}
where $d=L/N$ is the average inter-particle distance.
Detailed calculations are found in the SM. II. 
Note that our approximated solution exhibits a singularity at $\alpha=0$, since the dilute limit is broken in that case.

Comparing Eq. (\ref{Dominant channel momentum distribution}) with the two-body solution Eq. (\ref{E9}), one finds that only imaginary parts of momenta change, which are amplified by the number of particles with opposite spins.
This fact stems from the nature of resonant states.
For a spin-down particle, it deems each spin-up particle as a soft boundary, such that its length free of collision is roughly $L/N_\uparrow$.
Hence, the imaginary part of $k_{i,\downarrow}$ in Eq. (\ref{Dominant channel momentum distribution}) is amplified by a factor of $N_\uparrow$.
The case for a spin-up particle is in parallel. 

\begin{figure}[tp]
  \flushleft
  \includegraphics[scale=0.37]{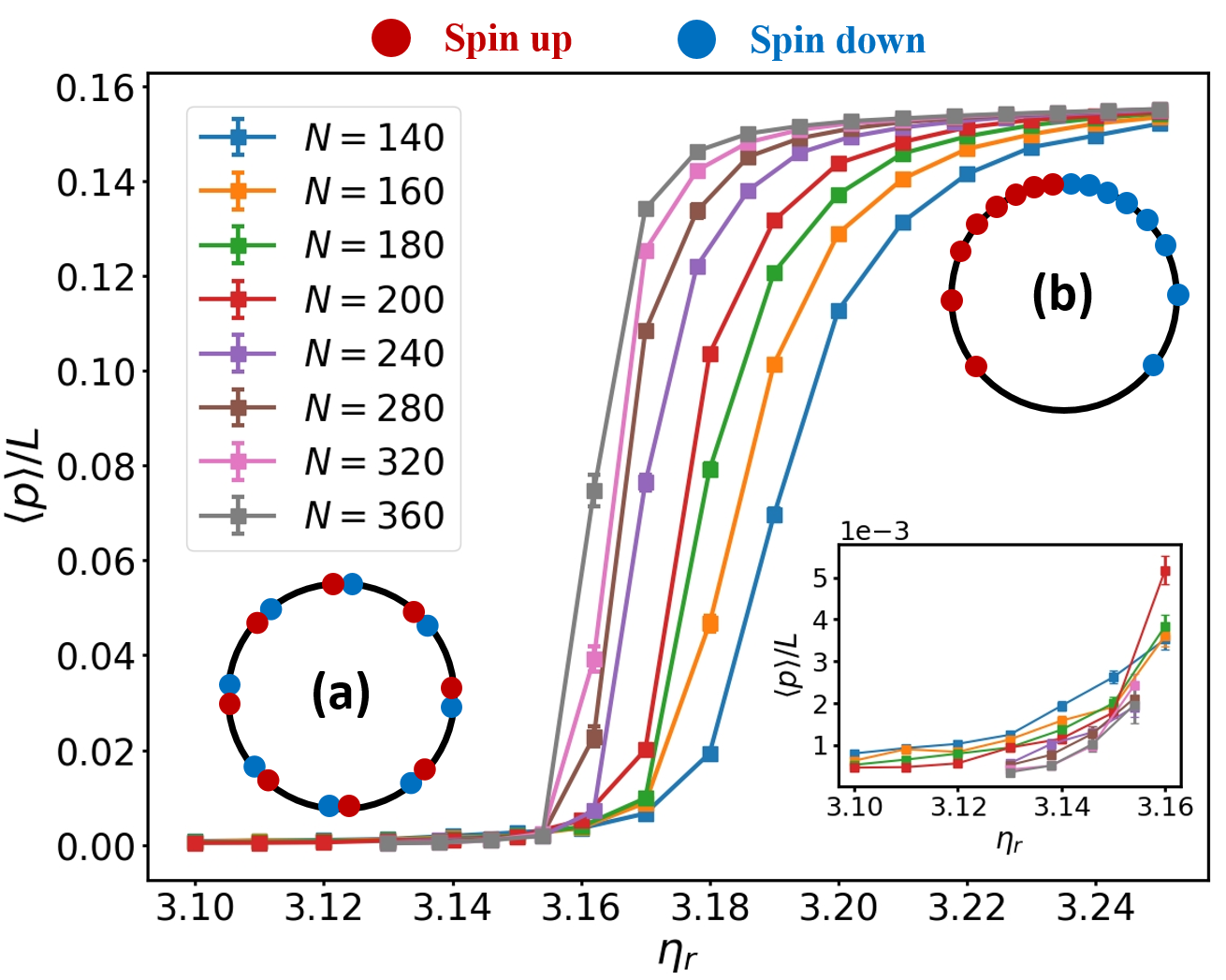}
  \caption{The Monte-Carlo simulations of the average `spin dipole' $\langle p \rangle$ at $L=100$ with different particle numbers.
  As $\eta_r = \ln(1+g/\alpha)$ increases, $\langle p \rangle/L$ jumps from zero to a finite value, indicating that the transition is of 1st order.
  ($a$) and ($b$) shows the configuration of the two phases.
  If $\alpha$ is fixed, the more localized ($b$) phase emerges only when $g$ is big enough, which implies that the many-body resonance is enhanced by the repulsive interaction.
  }
  \label{Figure of potential}
\end{figure}

{\em BA wavefunctions.}---
We denote $\varphi (x_{\downarrow}; x_{\uparrow})$ as an abbreviation to the following wavefunction,
\begin{equation*}
    \! \!    \varphi_{\underbrace{\downarrow \downarrow \! \cdots \! \downarrow}_{ N_\downarrow } \ 
    \underbrace{\uparrow \uparrow \! \cdots \! \uparrow}_{N_\uparrow} }  
    \left(x_{1}, \ x_{2} \cdots \! \! \ x_{N_\downarrow}; 
    \ x_{N_\downarrow + 1}, \ x_{N_\downarrow + 2} \cdots \! \! \ x_{N}\right). 
\end{equation*}
Other spin configurations of the BA wavefunction can be obtained by permutations according to Fermi statistics. 

In general, the BA wavefunction is very complex.
Fortunately, in the dilute limit, it can be factorized a similar way to that of the two-body case in Eq. (\ref{pair wavefunction}), 
\begin{equation}
\begin{aligned}
    \! \! \!
    \varphi (x_{\downarrow}; x_{\uparrow}) \! = \! 
    \text{det} (e^{\frac{i \chi_{i_1} x_{i_2} }{L} } ) 
    \text{det} (e^{\frac{i \chi_{j_1} x_{j_2} }{L} } )
    e^{-\frac{1}{2} \sum_{i j} \! W(x_{ij})},
    \label{Simplified Many-body wave function}
\end{aligned}
\end{equation}
in which two $\text{det}(\cdots)$ represent the Slater determinants of the plane-waves for spin-up and down particles respectively; 
the two-body Jastrow factor in Eq. (\ref{pair wavefunction}) is also generalized to the many-body case.
Here $1 \leq i,i_1,i_2 \leq N_\downarrow$ and $N_\downarrow+1 \leq j,j_1,j_2 \leq N$
are the coordinate indexes for spin-down and up particles respectively, 
and $x_{ij} = x_{i} -x_{j}$ represents the distance between them.
This wavefunction is similar to that of the Hubbard model at $U \rightarrow \infty$ \cite{ogata1990}, where the BA wavefunction is factorized into a Slater determinaint of spinless fermions and a BA wavefunction of the spin-1/2 chain. 

The above simplification is justified as follows.
As illustrated in Fig. \ref{Illustration of BA wavefunction}, the wavefunction is approximated to a summation of incident waves with different coordinate permutations.
In other words,
\begin{equation*}
\begin{aligned}
    \varphi (x_{\downarrow}; x_{\uparrow})&= 
    \text{det} (e^{\frac{i \chi_{i_1} x_{i_2} }{L} } ) 
    \text{det} (e^{\frac{i \chi_{j_1} x_{j_2} }{L} } )
    e^{-\sum_{ij} \frac{\eta_r x_{ij}}{L} } \\
    &\times \sum_{Q} \theta\left(x_{Q_1} > x_{Q_2} > \cdots > x_{Q_N} \right) A(Q),
\end{aligned}    
\end{equation*}
where $Q$ represents the permutation $x_{Q_1} > x_{Q_2} > \cdots > x_{Q_N}$ with $1 \leq Q_l \leq N$.
$\sum_Q$ denotes the sum over all permutations. 
These incident waves differ by a `phase shift', whose module is not $1$ due to the non-Hermitian SOC.
After switching a pair of neighboring particles with opposite spins, the amplitudes are changed by
\begin{equation}
    \frac{A\left(\cdots \uparrow\downarrow \cdots \right)}
    {A\left(\cdots  \downarrow \uparrow \cdots \right)} = 1+\frac{g}{\alpha}
    +O(\frac{\lambda_s}{d}),
\end{equation}
which is momentum independent
at the leading order.
As proved in SM. II, in this case the summation of step functions can be organized into 
\begin{equation*}
    \sum_{Q} \theta\left(x_{Q_1} > x_{Q_2} > \cdots > x_{Q_N} \right) A(Q) = 
    e^{-\eta_r \sum_{ij} \theta(x_j-x_i)}.
\end{equation*}
Further simplification yields the wavefunction Eq. (\ref{Simplified Many-body wave function}).

{\em Application.---} 
As an application of the above solution, we identify a phase transition in our system. 
Consider the case with an equal number of spin-up and spin-down particles, whose real parts of the momenta are $2\pi n_l/L$. 
Without loss of generality, assume the particle numbers of both components are odd.
The quantum numbers $n_j$ and $n_i$ for spin-up and down particles take the values of
\begin{equation*}
    -\frac{N-2}{4}, -\frac{N-6}{4}, \cdots
    -1, 0, 1, \cdots \frac{N-6}{4}, \frac{N-2}{4}.
\end{equation*}
In this case, the Slater determinant simplifies to \cite{gros1987antiferromagnetic}
\begin{equation*}
    \text{det} (e^{i \frac{\chi_{i_1} x_{i_2}}{L}}) = 
    \prod_{i_1<i_2}\left(
    2 i \sin 
    \frac{\pi \left(x_{i_1}-x_{i_2}\right)}{L}
    \right).
\end{equation*}

The probability distribution function $|\varphi (x_{\downarrow}; x_{\uparrow})|^2$ can be expressed as a thermodynamic distribution, similar to
the case of the Laughlin wave function 
\cite{laughlin1983anomalous}:
\begin{equation}
|\varphi (x_{\downarrow}; x_{\uparrow})|^2 = 
\rho (x_{\downarrow}; x_{\uparrow}) = e^{-\mathcal{H}}.
\label{eq:probility}
\end{equation}
The effective Hamiltonian $\mathcal{H}$ is
\begin{equation}
    \mathcal{H} = 
    \sum_{i_1<i_2}  V(x_{i_1, i_2}) +  
    \sum_{j_1<j_2}  V(x_{j_1, j_2}) + 
    \sum_{ij} W(x_{ij}),
\end{equation}
where $x_{i_1, i_2} = x_{i_1} - x_{i_2}$ and $x_{j_1, j_2} = x_{j_1} - x_{j_2}$.
$V (r) = -2 \ln \left| \sin \frac{\pi}{L} r \right|$ originates from the Pauli exclusion principle, which describes an effective repulsion between particles of identical spins;
$W$ brings an unidirectional attraction between opposite spins, with spin-up particles preferring the left side of spin-down ones.


If $V$ dominates, particles tend to uniformly distribute along the ring, with a weak pairing tendency between opposite spins to take the advantage of $W$, as depicted in Fig. \ref{Figure of potential} ($a$).
Conversely, if $W$ dominates, the system prefers phase separation, such that nearly all spin-up particles lie on the left of spin-down particles, making the configuration in Fig.~\ref{Figure of potential} ($b$) more favorable. 
The competition between $V$ and $W$ is investigated via Monte-Carlo simulations with the probability distribution Eq. (\ref{eq:probility}).
We consider the ``dipole" strength $p=\frac{1}{(N/2)^2} \sum_{ij} p_{ij}$, where $p_{ij}$ is the dipole between a spin-down and up particle located at $x_{i}$ and $x_{j}$ respectively,
\begin{equation}
p_{ij}=\left\{\begin{array}{l}
x_{ij}, ~~~~~~~~~~~~~~~~~ |x_{ij}| \leq L/2, \\
x_{ij} - \operatorname{sgn}(x_{ij}) L, \ |x_{ij}| > L/2.
\end{array}
\right.
\label{eq:dipole} 
\end{equation}
With this definition, $-L/2< p \leq L/2$. 
The thermodynamics average $\langle p \rangle$ is plotted in Fig. \ref{Figure of potential}.
As $\eta_r$ increases, $\langle p \rangle/L$ evolves from zero, which is consistent with Fig.~\ref{Figure of potential} ($a$), 
to a finite value, illustrated in Fig.~\ref{Figure of potential} ($b$).
The transition takes places at $\eta_r \approx 3.15$.
Note that $\eta_r$ measures the strength of resonance, which is enhanced by the repulsive interaction as shown in the two-body case.
Therefore, the emergence of the ($b$) phase demonstrates a many-body `skin effect' driven by the repulsive interaction, which is in stark contrast to the conventional NHSE \cite{Masahito2021DissipatHubbard, Pan2023BoseGas, WZhong2023SpinChain, Kattel2023SpinChain, CShu2024BoseHubbard}.

The abrupt change of the dipole strength $p$ at the transition point indicates a first-order phase transition.
Let us gain a better understanding by introducing an effective ``temperature" defined as
\begin{equation*}
    \rho_\beta = e^{-\beta \mathcal{H}}, 
\end{equation*}
in which $\beta=1$ corresponds to the situation described by Eq. (\ref{eq:probility}).
Consider the zero-temperature limit $\beta\rightarrow \infty$ such that the system freeze into the minimal energy configuration of $\mathcal{H}$. 
As an example, we examine the case of 4 particles shown in SM. III. 
The energy minima at small and large values of $\eta_r$ are calculated, which correspond to frozen configurations shown in 
Fig. \ref{Figure of potential} ($a$) and ($b$) respectively. 
The switch of minima occurs at $\eta_r=3.84$, roughly matching the transition point shown in Fig. \ref{Figure of potential}.



{\em Discussion and conclusion}.--- 
We present a BA solution to a 1D interacting spin-$\frac{1}{2}$ non-Hermitian system breaking the inversion symmetry. 
The interplay between non-Hermitian SOC and the repulsive interaction results in a novel many-body resonance state.
The complicated BA wavefunction is simplified in the dilute limit, which is cast into the Slater-Jastrow form.
Its amplitude square is mapped into a thermodynamic distribution, exhibiting the competition between Pauli's exclusion among fermions of the same component and resonances between fermions of different components.
The former brings a repulsion and the latter generates an unidirectional attraction. 
This competition leads to a transition from a uniform configuration with weak pairing tendency to a phase separation, showing that the many-body resonance is enhanced by the repulsive interaction.


\begin{acknowledgments}
We are grateful to K. Yang, W. Yang, J. D. Wu and C. H. Ke for valuable discussions. 
Z. Y. is supported by the National Key Research and Development Program of China (Grant No. 2023YFA1407500), the National Natural Science Foundation of China (12322405, 12104450, 12047503), and the Fundamental Research Funds for the Central Universities (20720230011).
C.W. is supported by the National Natural Science Foundation of China under the Grants No. 12234016 and No. 12174317.
This work has been supported by the New Cornerstone Science Foundation.
\end{acknowledgments}

\nocite{*}

\bibliography{Citation}

@article{gros1987antiferromagnetic,
  title = {Antiferromagnetic correlations in almost-localized Fermi liquids},
  author = {Gros, C. and Joynt, R. and Rice, T. M.},
  journal = {Phys. Rev. B},
  volume = {36},
  issue = {1},
  pages = {381--393},
  numpages = {0},
  year = {1987},
  month = {Jul},
  publisher = {American Physical Society},
  doi = {10.1103/PhysRevB.36.381},
  url = {https://link.aps.org/doi/10.1103/PhysRevB.36.381}
}

@article{laughlin1983anomalous,
  title = {Anomalous Quantum Hall Effect: An Incompressible Quantum Fluid with Fractionally Charged Excitations},
  author = {Laughlin, R. B.},
  journal = {Phys. Rev. Lett.},
  volume = {50},
  issue = {18},
  pages = {1395--1398},
  numpages = {0},
  year = {1983},
  month = {May},
  publisher = {American Physical Society},
  doi = {10.1103/PhysRevLett.50.1395},
  url = {https://link.aps.org/doi/10.1103/PhysRevLett.50.1395}
}

@article{lee2016anomalous,
  title = {Anomalous Edge State in a Non-Hermitian Lattice},
  author = {Lee, Tony E.},
  journal = {Phys. Rev. Lett.},
  volume = {116},
  issue = {13},
  pages = {133903},
  numpages = {5},
  year = {2016},
  month = {Apr},
  publisher = {American Physical Society},
  doi = {10.1103/PhysRevLett.116.133903},
  url = {https://link.aps.org/doi/10.1103/PhysRevLett.116.133903}
}

@article{martinez2018non,
  title = {Non-Hermitian robust edge states in one dimension: Anomalous localization and eigenspace condensation at exceptional points},
  author = {Martinez Alvarez, V. M. and Barrios Vargas, J. E. and Foa Torres, L. E. F.},
  journal = {Phys. Rev. B},
  volume = {97},
  issue = {12},
  pages = {121401},
  numpages = {6},
  year = {2018},
  month = {Mar},
  publisher = {American Physical Society},
  doi = {10.1103/PhysRevB.97.121401},
  url = {https://link.aps.org/doi/10.1103/PhysRevB.97.121401}
}

@article{yao2018edge,
  title = {Edge States and Topological Invariants of Non-Hermitian Systems},
  author = {Yao, Shunyu and Wang, Zhong},
  journal = {Phys. Rev. Lett.},
  volume = {121},
  issue = {8},
  pages = {086803},
  numpages = {8},
  year = {2018},
  month = {Aug},
  publisher = {American Physical Society},
  doi = {10.1103/PhysRevLett.121.086803},
  url = {https://link.aps.org/doi/10.1103/PhysRevLett.121.086803}
}

@article{lee2019hybrid,
  title = {Hybrid Higher-Order Skin-Topological Modes in Nonreciprocal Systems},
  author = {Lee, Ching Hua and Li, Linhu and Gong, Jiangbin},
  journal = {Phys. Rev. Lett.},
  volume = {123},
  issue = {1},
  pages = {016805},
  numpages = {6},
  year = {2019},
  month = {Jul},
  publisher = {American Physical Society},
  doi = {10.1103/PhysRevLett.123.016805},
  url = {https://link.aps.org/doi/10.1103/PhysRevLett.123.016805}
}

@article{lee2019anatomy,
  title = {Anatomy of skin modes and topology in non-Hermitian systems},
  author = {Lee, Ching Hua and Thomale, Ronny},
  journal = {Phys. Rev. B},
  volume = {99},
  issue = {20},
  pages = {201103},
  numpages = {5},
  year = {2019},
  month = {May},
  publisher = {American Physical Society},
  doi = {10.1103/PhysRevB.99.201103},
  url = {https://link.aps.org/doi/10.1103/PhysRevB.99.201103}
}

@article{borgnia2020non,
  title = {Non-Hermitian Boundary Modes and Topology},
  author = {Borgnia, Dan S. and Kruchkov, Alex Jura and Slager, Robert-Jan},
  journal = {Phys. Rev. Lett.},
  volume = {124},
  issue = {5},
  pages = {056802},
  numpages = {6},
  year = {2020},
  month = {Feb},
  publisher = {American Physical Society},
  doi = {10.1103/PhysRevLett.124.056802},
  url = {https://link.aps.org/doi/10.1103/PhysRevLett.124.056802}
}

@article{SuPeng2020Edge,
  title = {Defective edge states and number-anomalous bulk-boundary correspondence in non-Hermitian topological systems},
  author = {Wang, Xiao-Ran and Guo, Cui-Xian and Kou, Su-Peng},
  journal = {Phys. Rev. B},
  volume = {101},
  issue = {12},
  pages = {121116},
  numpages = {5},
  year = {2020},
  month = {Mar},
  publisher = {American Physical Society},
  doi = {10.1103/PhysRevB.101.121116},
  url = {https://link.aps.org/doi/10.1103/PhysRevB.101.121116}
}

@article{SuPeng2024ChiralSkin,
  title = {Non-Hermitian chiral skin effect},
  author = {Ma, Xin-Ran and Cao, Kui and Wang, Xiao-Ran and Wei, Zheng and Du, Qian and Kou, Su-Peng},
  journal = {Phys. Rev. Res.},
  volume = {6},
  issue = {1},
  pages = {013213},
  numpages = {17},
  year = {2024},
  month = {Feb},
  publisher = {American Physical Society},
  doi = {10.1103/PhysRevResearch.6.013213},
  url = {https://link.aps.org/doi/10.1103/PhysRevResearch.6.013213}
}

@article{Lieb1963BoseGasI,
  title = {Exact Analysis of an Interacting Bose Gas. I. The General Solution and the Ground State},
  author = {Lieb, Elliott H. and Liniger, Werner},
  journal = {Phys. Rev.},
  volume = {130},
  issue = {4},
  pages = {1605--1616},
  numpages = {0},
  year = {1963},
  month = {May},
  publisher = {American Physical Society},
  doi = {10.1103/PhysRev.130.1605},
  url = {https://link.aps.org/doi/10.1103/PhysRev.130.1605}
}

@article{Lieb1963BoseGasII,
  title = {Exact Analysis of an Interacting Bose Gas. II. The Excitation Spectrum},
  author = {Lieb, Elliott H.},
  journal = {Phys. Rev.},
  volume = {130},
  issue = {4},
  pages = {1616--1624},
  numpages = {0},
  year = {1963},
  month = {May},
  publisher = {American Physical Society},
  doi = {10.1103/PhysRev.130.1616},
  url = {https://link.aps.org/doi/10.1103/PhysRev.130.1616}
}

@article{Yang1967FermiGas,
  title = {Some Exact Results for the Many-Body Problem in one Dimension with Repulsive Delta-Function Interaction},
  author = {Yang, C. N.},
  journal = {Phys. Rev. Lett.},
  volume = {19},
  issue = {23},
  pages = {1312--1315},
  numpages = {0},
  year = {1967},
  month = {Dec},
  publisher = {American Physical Society},
  doi = {10.1103/PhysRevLett.19.1312},
  url = {https://link.aps.org/doi/10.1103/PhysRevLett.19.1312}
}

@article{GAUDIN1967FermiGas,
title = {Un systeme a une dimension de fermions en interaction},
journal = {Physics Letters A},
volume = {24},
number = {1},
pages = {55-56},
year = {1967},
issn = {0375-9601},
doi = {https://doi.org/10.1016/0375-9601(67)90193-4},
url = {https://www.sciencedirect.com/science/article/pii/0375960167901934},
author = {M. Gaudin}
}

@article{Lieb1968Hubbard,
  title = {Absence of Mott Transition in an Exact Solution of the Short-Range, One-Band Model in One Dimension},
  author = {Lieb, Elliott H. and Wu, F. Y.},
  journal = {Phys. Rev. Lett.},
  volume = {20},
  issue = {25},
  pages = {1445--1448},
  numpages = {0},
  year = {1968},
  month = {Jun},
  publisher = {American Physical Society},
  doi = {10.1103/PhysRevLett.20.1445},
  url = {https://link.aps.org/doi/10.1103/PhysRevLett.20.1445}
}

@book{korepin1997quantum,
  title={Quantum inverse scattering method and correlation functions},
  author={Korepin, Vladimir E and Korepin, Vladimir E and Bogoliubov, NM and Izergin, AG},
  volume={3},
  year={1997},
  publisher={Cambridge university press}
}

@book{wang2015off,
  title={Off-diagonal Bethe ansatz for exactly solvable models},
  author={Wang, Yupeng and Yang, Wen-Li and Cao, Junpeng and Shi, Kangjie},
  year={2015},
  publisher={Springer}
}

@article{Emil2018BulkBoundary,
  title = {Biorthogonal Bulk-Boundary Correspondence in Non-Hermitian Systems},
  author = {Kunst, Flore K. and Edvardsson, Elisabet and Budich, Jan Carl and Bergholtz, Emil J.},
  journal = {Phys. Rev. Lett.},
  volume = {121},
  issue = {2},
  pages = {026808},
  numpages = {6},
  year = {2018},
  month = {Jul},
  publisher = {American Physical Society},
  doi = {10.1103/PhysRevLett.121.026808},
  url = {https://link.aps.org/doi/10.1103/PhysRevLett.121.026808}
}

@article{Yao2018ChernBand,
  title = {Non-Hermitian Chern Bands},
  author = {Yao, Shunyu and Song, Fei and Wang, Zhong},
  journal = {Phys. Rev. Lett.},
  volume = {121},
  issue = {13},
  pages = {136802},
  numpages = {8},
  year = {2018},
  month = {Sep},
  publisher = {American Physical Society},
  doi = {10.1103/PhysRevLett.121.136802},
  url = {https://link.aps.org/doi/10.1103/PhysRevLett.121.136802}
}

@article{Song2019OpenSys,
  title = {Non-Hermitian Skin Effect and Chiral Damping in Open Quantum Systems},
  author = {Song, Fei and Yao, Shunyu and Wang, Zhong},
  journal = {Phys. Rev. Lett.},
  volume = {123},
  issue = {17},
  pages = {170401},
  numpages = {8},
  year = {2019},
  month = {Oct},
  publisher = {American Physical Society},
  doi = {10.1103/PhysRevLett.123.170401},
  url = {https://link.aps.org/doi/10.1103/PhysRevLett.123.170401}
}

@article{Murakami2019Band,
  title = {Non-Bloch Band Theory of Non-Hermitian Systems},
  author = {Yokomizo, Kazuki and Murakami, Shuichi},
  journal = {Phys. Rev. Lett.},
  volume = {123},
  issue = {6},
  pages = {066404},
  numpages = {6},
  year = {2019},
  month = {Aug},
  publisher = {American Physical Society},
  doi = {10.1103/PhysRevLett.123.066404},
  url = {https://link.aps.org/doi/10.1103/PhysRevLett.123.066404}
}

@article{Yang2020BulkBoundary,
  title = {Non-Hermitian Bulk-Boundary Correspondence and Auxiliary Generalized Brillouin Zone Theory},
  author = {Yang, Zhesen and Zhang, Kai and Fang, Chen and Hu, Jiangping},
  journal = {Phys. Rev. Lett.},
  volume = {125},
  issue = {22},
  pages = {226402},
  numpages = {6},
  year = {2020},
  month = {Nov},
  publisher = {American Physical Society},
  doi = {10.1103/PhysRevLett.125.226402},
  url = {https://link.aps.org/doi/10.1103/PhysRevLett.125.226402}
}

@article{Yang2020Correspondence,
  title = {Correspondence between Winding Numbers and Skin Modes in Non-Hermitian Systems},
  author = {Zhang, Kai and Yang, Zhesen and Fang, Chen},
  journal = {Phys. Rev. Lett.},
  volume = {125},
  issue = {12},
  pages = {126402},
  numpages = {6},
  year = {2020},
  month = {Sep},
  publisher = {American Physical Society},
  doi = {10.1103/PhysRevLett.125.126402},
  url = {https://link.aps.org/doi/10.1103/PhysRevLett.125.126402}
}

@article{brandenbourger2019non,
  title={Non-reciprocal robotic metamaterials},
  author={Brandenbourger, Martin and Locsin, Xander and Lerner, Edan and Coulais, Corentin},
  journal={Nature communications},
  volume={10},
  number={1},
  pages={4608},
  year={2019},
  publisher={Nature Publishing Group UK London},
  doi = {10.1038/s41467-019-12599-3}, 
  url = {https://www.nature.com/articles/s41467-019-12599-3}
}

@article{xiao2020non,
  title={Non-Hermitian bulk--boundary correspondence in quantum dynamics},
  author={Xiao, Lei and Deng, Tianshu and Wang, Kunkun and Zhu, Gaoyan and Wang, Zhong and Yi, Wei and Xue, Peng},
  journal={Nature Physics},
  volume={16},
  number={7},
  pages={761--766},
  year={2020},
  publisher={Nature Publishing Group UK London},
  doi = {10.1038/s41567-020-0836-6},
  url = {https://www.nature.com/articles/s41567-020-0836-6}
}

@article{ghatak2020observation,
  title={Observation of non-Hermitian topology and its bulk--edge correspondence in an active mechanical metamaterial},
  author={Ghatak, Ananya and Brandenbourger, Martin and Van Wezel, Jasper and Coulais, Corentin},
  journal={Proceedings of the National Academy of Sciences},
  volume={117},
  number={47},
  pages={29561--29568},
  year={2020},
  publisher={National Acad Sciences},
  doi = {10.1073/pnas.2010580117},
  url = {https://www.pnas.org/doi/abs/10.1073/pnas.2010580117}
}

@article{hofmann2020reciprocal,
  title = {Reciprocal skin effect and its realization in a topolectrical circuit},
  author = {Hofmann, Tobias and Helbig, Tobias and Schindler, Frank and Salgo, Nora and Brzezi\ifmmode \acute{n}\else \'{n}\fi{}ska, Marta and Greiter, Martin and Kiessling, Tobias and Wolf, David and Vollhardt, Achim and Kaba\ifmmode \check{s}\else \v{s}\fi{}i, Anton and Lee, Ching Hua and Bilu\ifmmode \check{s}\else \v{s}\fi{}i\ifmmode \acute{c}\else \'{c}\fi{}, Ante and Thomale, Ronny and Neupert, Titus},
  journal = {Phys. Rev. Res.},
  volume = {2},
  issue = {2},
  pages = {023265},
  numpages = {11},
  year = {2020},
  month = {Jun},
  publisher = {American Physical Society},
  doi = {10.1103/PhysRevResearch.2.023265},
  url = {https://link.aps.org/doi/10.1103/PhysRevResearch.2.023265}
}

@article{Zhang2021Circuit,
    author = {Shuo Liu  and Ruiwen Shao  and Shaojie Ma  and Lei Zhang  and Oubo You  and Haotian Wu  and Yuan Jiang Xiang  and Tie Jun Cui  and Shuang Zhang},
    title = {Non-Hermitian Skin Effect in a Non-Hermitian Electrical Circuit},
    journal = {Research},
    volume = {2021},
    number = {},
    pages = {},
    year = {2021},
    doi = {10.34133/2021/5608038},
    URL = {https://spj.science.org/doi/abs/10.34133/2021/5608038},
}

@article{zhang2021acoustic,
  title = {Acoustic non-Hermitian skin effect from twisted winding topology},
  author = {Zhang, Li and Yang, Yihao and Ge, Yong and Guan, Yi-Jun and Chen, Qiaolu and Yan, Qinghui and Chen, Fujia and Xi, Rui and Li, Yuanzhen and Jia, Ding and others},
  journal = {Nature communications},
  volume = {12},
  number = {1},
  pages = {6297},
  year = {2021},
  doi = {10.1038/s41467-021-26619-8},
  url = {https://www.nature.com/articles/s41467-021-26619-8},
  publisher={Nature Publishing Group UK London}
}

@article{Yanbo2022ColdAtom,
  title = {Dynamic Signatures of Non-Hermitian Skin Effect and Topology in Ultracold Atoms},
  author = {Liang, Qian and Xie, Dizhou and Dong, Zhaoli and Li, Haowei and Li, Hang and Gadway, Bryce and Yi, Wei and Yan, Bo},
  journal = {Phys. Rev. Lett.},
  volume = {129},
  issue = {7},
  pages = {070401},
  numpages = {6},
  year = {2022},
  month = {Aug},
  publisher = {American Physical Society},
  doi = {10.1103/PhysRevLett.129.070401},
  url = {https://link.aps.org/doi/10.1103/PhysRevLett.129.070401}
}

@article{Masahito2021DissipatHubbard,
  title = {Exact Liouvillian Spectrum of a One-Dimensional Dissipative Hubbard Model},
  author = {Nakagawa, Masaya and Kawakami, Norio and Ueda, Masahito},
  journal = {Phys. Rev. Lett.},
  volume = {126},
  issue = {11},
  pages = {110404},
  numpages = {6},
  year = {2021},
  month = {Mar},
  publisher = {American Physical Society},
  doi = {10.1103/PhysRevLett.126.110404},
  url = {https://link.aps.org/doi/10.1103/PhysRevLett.126.110404}
}

@article{Kattel2023SpinChain,
doi = {10.1088/1751-8121/ace56e},
url = {https://dx.doi.org/10.1088/1751-8121/ace56e},
year = {2023},
month = {jul},
publisher = {IOP Publishing},
volume = {56},
number = {32},
pages = {325001},
author = {Pradip Kattel and Parameshwar R Pasnoori and Natan Andrei},
title = {Exact solution of a non-Hermitian PT-symmetric spin chain},
journal = {Journal of Physics A: Mathematical and Theoretical}
}

@article{Pan2023BoseGas,
  title = {Non-Hermitian skin effect in a one-dimensional interacting Bose gas},
  author = {Mao, Liang and Hao, Yajiang and Pan, Lei},
  journal = {Phys. Rev. A},
  volume = {107},
  issue = {4},
  pages = {043315},
  numpages = {9},
  year = {2023},
  month = {Apr},
  publisher = {American Physical Society},
  doi = {10.1103/PhysRevA.107.043315},
  url = {https://link.aps.org/doi/10.1103/PhysRevA.107.043315}
}

@Article{WZhong2023SpinChain,
	title={{Scale-free non-Hermitian skin effect in a boundary-dissipated spin chain}},
	author={He-Ran Wang and Bo Li and Fei Song and Zhong Wang},
	journal={SciPost Phys.},
	volume={15},
	pages={191},
	year={2023},
	publisher={SciPost},
	doi={10.21468/SciPostPhys.15.5.191},
	url={https://scipost.org/10.21468/SciPostPhys.15.5.191},
}

@article{CShu2024BoseHubbard,
  title = {Exact Solution of the Bose-Hubbard Model with Unidirectional Hopping},
  author = {Zheng, Mingchen and Qiao, Yi and Wang, Yupeng and Cao, Junpeng and Chen, Shu},
  journal = {Phys. Rev. Lett.},
  volume = {132},
  issue = {8},
  pages = {086502},
  numpages = {7},
  year = {2024},
  month = {Feb},
  publisher = {American Physical Society},
  doi = {10.1103/PhysRevLett.132.086502},
  url = {https://link.aps.org/doi/10.1103/PhysRevLett.132.086502}
}

@article{Wu2019String, 
  title = {One-dimensional quantum spin dynamics of Bethe string states},
  author = {Yang, Wang and Wu, Jianda and Xu, Shenglong and Wang, Zhe and Wu, Congjun},
  journal = {Phys. Rev. B},
  volume = {100},
  issue = {18},
  pages = {184406},
  numpages = {14},
  year = {2019},
  month = {Nov},
  publisher = {American Physical Society},
  doi = {10.1103/PhysRevB.100.184406},
  url = {https://link.aps.org/doi/10.1103/PhysRevB.100.184406}
}

@article{Wu2018ExString,
  title={Experimental observation of Bethe strings},
  author={Wang, Zhe and Wu, Jianda and Yang, Wang and Bera, Anup Kumar and Kamenskyi, Dmytro and Islam, ATM Nazmul and Xu, Shenglong and Law, Joseph Matthew and Lake, Bella and Wu, Congjun and others},
  journal={Nature},
  volume={554},
  number={7691},
  pages={219--223},
  year={2018},
  doi = {https://doi.org/10.1038/nature25466},
  url = {https://www.nature.com/articles/nature25466},
  publisher={Nature Publishing Group UK London}
}

@book{takahashi,
  title={Thermodynamics of One-Dimensional Solvable Models},
  author={Takahashi, M.},
  isbn={9780521551434},
  lccn={98034997},
  url={https://books.google.com/books?id=kX1FAwEACAAJ},
  doi={https://doi.org/10.1017/CBO9780511524332},
  year={1999},
  publisher={Cambridge University Press}
}

@article{Jastrow,
  title = {Many-Body Problem with Strong Forces},
  author = {Jastrow, Robert},
  journal = {Phys. Rev.},
  volume = {98},
  issue = {5},
  pages = {1479--1484},
  numpages = {0},
  year = {1955},
  month = {Jun},
  publisher = {American Physical Society},
  doi = {10.1103/PhysRev.98.1479},
  url = {https://link.aps.org/doi/10.1103/PhysRev.98.1479}
}

@article{ogata1990,
  title = {Bethe-ansatz wave function, momentum distribution, and spin correlation in the one-dimensional strongly correlated Hubbard model},
  author = {Ogata, Masao and Shiba, Hiroyuki},
  journal = {Phys. Rev. B},
  volume = {41},
  issue = {4},
  pages = {2326--2338},
  numpages = {0},
  year = {1990},
  month = {Feb},
  publisher = {American Physical Society},
  doi = {10.1103/PhysRevB.41.2326},
  url = {https://link.aps.org/doi/10.1103/PhysRevB.41.2326}
}

@article{Yang-2020-Spin-Dependent-NHSE,
  title = {Non-Hermitian Skin Modes Induced by On-Site Dissipations and Chiral Tunneling Effect},
  author = {Yi, Yifei and Yang, Zhesen},
  journal = {Phys. Rev. Lett.},
  volume = {125},
  issue = {18},
  pages = {186802},
  numpages = {7},
  year = {2020},
  month = {Oct},
  publisher = {American Physical Society},
  doi = {10.1103/PhysRevLett.125.186802},
  url = {https://link.aps.org/doi/10.1103/PhysRevLett.125.186802}
}

@article{li2021impurity,
  title={Impurity induced scale-free localization},
  author={Li, Linhu and Lee, Ching Hua and Gong, Jiangbin},
  journal={Communications Physics},
  volume={4},
  number={1},
  pages={42},
  year={2021},
  publisher={Nature Publishing Group UK London},
  url={https://www.nature.com/articles/s42005-021-00547-x}
}

@article{li2020critical,
  title={Critical non-Hermitian skin effect},
  author={Li, Linhu and Lee, Ching Hua and Mu, Sen and Gong, Jiangbin},
  journal={Nature communications},
  volume={11},
  number={1},
  pages={5491},
  year={2020},
  publisher={Nature Publishing Group UK London},
  url={https://www.nature.com/articles/s41467-020-18917-4}
}

@Article{SciPostPhys.8.3.044,
    title={{Yang-Baxter integrable Lindblad equations}},
    author={Aleksandra A. Ziolkowska and Fabian H.L. Essler},
    journal={SciPost Phys.},
    volume={8},
    pages={044},
    year={2020},
    publisher={SciPost},
    doi={10.21468/SciPostPhys.8.3.044},
    url={https://scipost.org/10.21468/SciPostPhys.8.3.044},
}

@article{New.J.Phys.22_123040,
    doi = {10.1088/1367-2630/abd124},
    url = {https://dx.doi.org/10.1088/1367-2630/abd124},
    year = {2020},
    month = {dec},
    publisher = {IOP Publishing},
    volume = {22},
    number = {12},
    pages = {123040},
    author = {Buča, Berislav and Booker, Cameron and Medenjak, Marko and Jaksch, Dieter},
    title = {Bethe ansatz approach for dissipation: exact solutions of quantum many-body dynamics under loss},
    journal = {New Journal of Physics},
}

@article{PhysRevResearch.6.L032067,
  title = {Liouvillian skin effects and fragmented condensates in an integrable dissipative Bose-Hubbard model},
  author = {Ekman, Christopher and Bergholtz, Emil J.},
  journal = {Phys. Rev. Res.},
  volume = {6},
  issue = {3},
  pages = {L032067},
  numpages = {7},
  year = {2024},
  month = {Sep},
  publisher = {American Physical Society},
  doi = {10.1103/PhysRevResearch.6.L032067},
  url ={https://link.aps.org/doi/10.1103/PhysRevResearch.6.L032067}
}

@article{girardeau1960relationship,
  title={Relationship between systems of impenetrable bosons and fermions in one dimension},
  author={Girardeau, Marvin},
  journal={Journal of Mathematical Physics},
  volume={1},
  number={6},
  pages={516--523},
  year={1960},
  publisher={American Institute of Physics}
}

@article{buvca2020bethe,
  title={Bethe ansatz approach for dissipation: exact solutions of quantum many-body dynamics under loss},
  author={Bu{\v{c}}a, Berislav and Booker, Cameron and Medenjak, Marko and Jaksch, Dieter},
  journal={New Journal of Physics},
  volume={22},
  number={12},
  pages={123040},
  year={2020},
  publisher={IOP Publishing}
}

\section{End Matter}
As for experimental realizations, a 1D lattice Hamiltonian was proposed in  \cite{Yang-2020-Spin-Dependent-NHSE}, in which the spin-dependent coupling between two sublattices transfers the onsite loss into the non-Hermitian SOC studied in our work.
Upon open boundary conditions, it exhibits a spin-dependent NHSE, which is a signature of the non-Hermitian SOC (More details are presented in SM IV).
It is expected that upon turning on repulsions and switching to the periodical boundary condition, such a system will exhibit the many-body resonance effect qualitatively similar to what we have studied.
Due to the sublattice structure, such a model is no longer integrable. 
Further numerical investigations are deferred to a later publication. 

For realistic experimental systems with pure loss, the dynamics should be governed by the Lindblad equation, which is more challenging to solve compared to the non-Hermitian Hamiltonian. 
Full treatment of this problem will be deferred to future publication. 
Nevertheless, we believe that the resonance effect derived from the many-body wavefunction Eq. (16) is robust enough even under particle loss. 
The key point is that the many-body resonance we identify is not a property of a single many-body eigenstate but is instead shared by all eigenstates. 
The robustness of the many-body resonance is also evident within the Lindbladian formalism. 
For purely lossy systems, the eigenmodes of the Lindblad super-operator are linear combinations of the basis constructed by the eigenstates of the corresponding non-Hermitian Hamiltonian. 
Supposition of states within the same particle-number sector preserve the resonance effect, since all the eigenstates share a common Jastrow factor. 
As shown in Eq. (16), the Jastrow factor is only determined by the numbers of spin-up and spin-down particles and is independent of the Bethe Ansatz quantum numbers. 
As the particle number decays over time, it should exhibit a distribution at each time. 
Because states in different particle-number sectors do not have spatial interference, the resonance effect is dominated by the most probable sectors of particle numbers. 
Therefore, the resonance effect should persist during particle loss, with quantitative modifications determined by the level particle-number fluctuations.

\clearpage

\begin{center}
\onecolumngrid
{\large\bfseries Supplemental Material}
\vspace{1em}
\end{center}

\twocolumngrid
\section{I. Exact solution of two-body problem} \label{Exact solution of two-body problem}

Quantization of $k_r$ is obtained via the following equation:
\begin{align*}
&\frac{\beta}{g}\Big( (-1)^{n_K} \operatorname{cosh} m \alpha L - \operatorname{cosh}m \beta L \Big) 
=\operatorname{sinh}m \beta L,
\end{align*}
with $\beta = \alpha + ik_r/m$.
Let us consider its solution in the case of $L \gg \lambda_s = (m \alpha)^{-1}$.
To balance the magnitudes of l.h.s and r.h.s, the real part of $\beta$ should be of same order with $\alpha$.
This simplifies the equation to
\begin{equation*}
    \frac{\beta}{g}\left( (-1)^{n_K} e^{-i k_r L} - 1 \right) = 1.
\end{equation*}
We consider those solutions with $k_r \sim 1/L$, in which case $\beta \approx \alpha$ in the leading order. 
$k_r$ can then be solved as
\begin{equation} 
    k_r = \frac{(2n_r+n_K)\pi}{L} + i\frac{\ln \left(1+g/\alpha \right)}{L} = 
    \frac{\chi_r + i\eta_r}{L}.
    \label{Definition of kr}
\end{equation}
In the limit $g \sim \frac{1}{mL}e^{L/\lambda_s}$, one finds $\Im k_r \sim \lambda_s^{-1}$, which is consistent with the solution under the OBC. 
As $g=0$, the free particle solution is recovered with $\Im k_r = 0$.

\begin{figure*}[tp]
    \centering
    \includegraphics[scale=0.2]{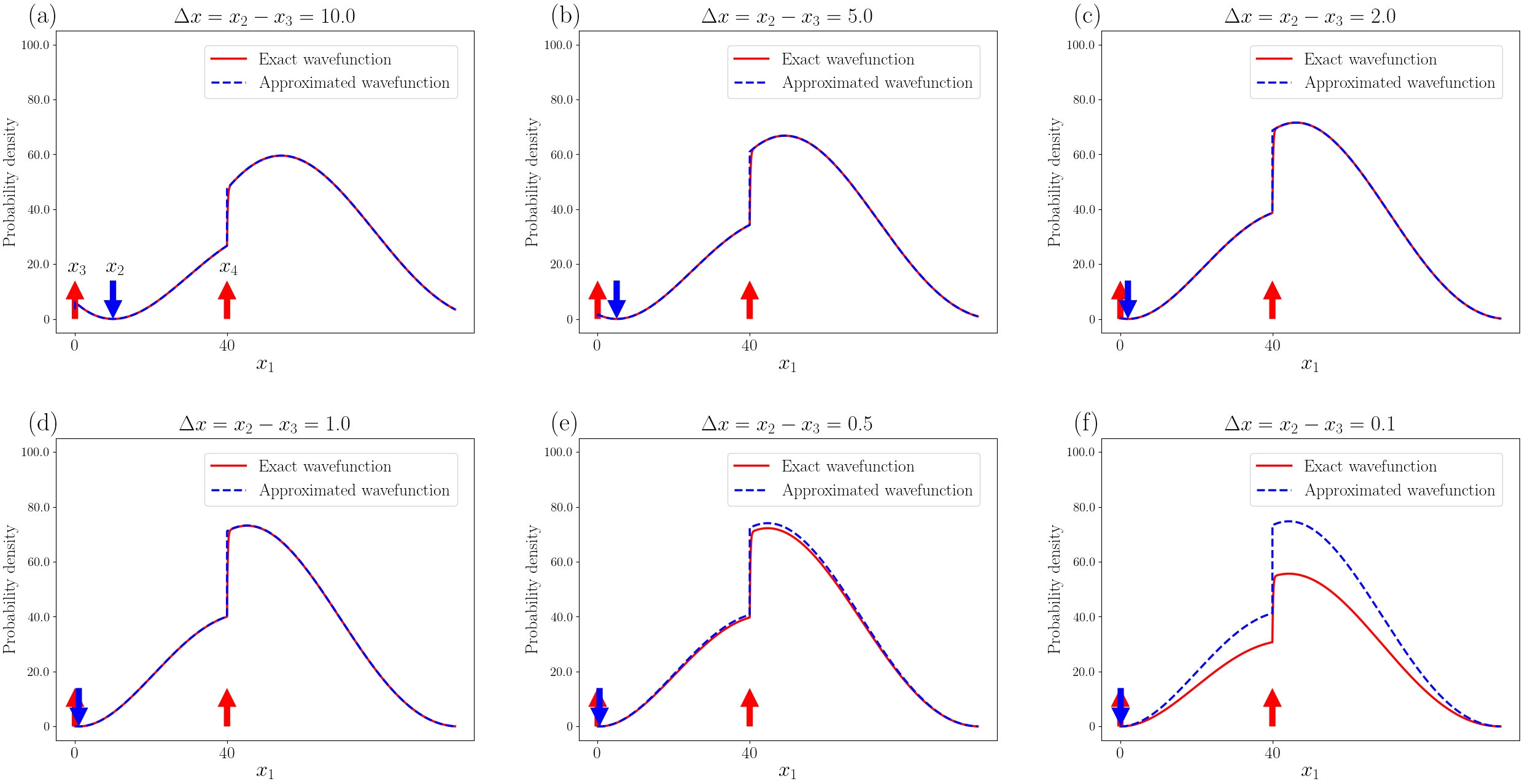}
    \caption{The probability distribution of four-particle wavefunction with $x_2, x_3, x_4$
    fixed. From (a) to (f), we decrease the value of $x_2$, with $x_3=0$, $x_4=40$, $\Delta x = x_2 - x_3$. 
    Here we take $\chi_{1, 2}=\chi_{3, 4}=\pm 2\pi/L$, $m=1$, $g=1$, $\alpha=3$, $L=100$.
    The normalization of the original wavefunction and the approximated wavefunction differs by
    a ratio 0.00765. As a comparison, $\lambda_s/L = 0.00333$.
    }
    \label{4-particle wavefunction}
\end{figure*}

\section{II. Solution to BA equation} \label{Solution to Bethe-ansatz equation}
We decouple the spin and momentum by applying a similar transformation $\hat{V}=\text{exp}(-m\alpha\sum_l \sigma^z_l x_l)$. 
The wavefunction is converted to $\overline{\psi}_{\sigma}(x_l) = \hat{V}\psi_{\sigma}(x_l)$, where $\sigma$ denotes the spin components. Transformed Hamiltonian $\hat{\overline{H}} = \hat{V} \hat{H} \hat{V}^{-1}$ is
\begin{equation*}
    \hat{\overline{H}} = \sum_l -\frac{\nabla_l^2}{2 m} + 
    \sum_{\langle l l' \rangle} 2g \delta\left(x_l-x_{l'}\right).
\end{equation*}
This transformation also twists the PBC to
\begin{equation*}
    \overline{\psi}_{\sigma}(x_l) = 
    e^{m\alpha L \sigma^z_l} \ \overline{\psi}_{\sigma}(x_l + L).
\end{equation*}
With this setup, the many-body problem can be solved using the BA method.
The Bethe-type wave function is defined as
\begin{equation}
\begin{aligned}
    \overline{\psi}_\sigma\left(x_l\right)= 
    &\sum_{Q, P} \theta\left(x_{Q_1} > x_{Q_2} > \cdots > x_{Q_N} \right) \\
    &\times A_\sigma(Q, P) \ \exp \left(i \sum_l \overline{k}_{P_l} x_{Q_l}\right),
    \label{Bare BA wavefunction}
\end{aligned}
\end{equation}
where $Q$ and $P$ represents permutations of the coordinates $\{ x_l \}$ and momenta $\{ \overline{k}_l \}$ respectively.
$\sum_{Q, P}$ denotes a sum over all permutations. 
For eigenstates with $N_\downarrow$ and $N_\uparrow = N - N_\downarrow$ spin-down and spin-up particles, the corresponding BA equations are
\begin{equation}
    \begin{aligned}
    &\prod_{i=1}^{N_\downarrow} \!
    \frac{\Lambda_{i}-\overline{k}_{l}-i m g}
    {\Lambda_{i}-\overline{k}_{l}+i m g} \! = \!
    e^{i \overline{k}_{l} L} e^{m \alpha L}, \\
    &\prod_{l=1}^{N} \!
    \frac{\Lambda_{i}-\overline{k}_{l}-i m g}
    {\Lambda_{i}-\overline{k}_{l}+i m g} \! = \! - 
    e^{2 m \alpha L} \prod_{i'=1}^{N_\downarrow} \!
    \frac{\Lambda_{i}-\Lambda_{i'}-2 i m g}
    {\Lambda_{i}-\Lambda_{i'}+2 i m g},
    \label{BA equation}
    \end{aligned}
\end{equation}
or equivalently
\begin{equation}
    \begin{aligned}
    &\prod_{j=1}^{N_\uparrow} \!
    \frac{\Gamma_{j} \! - \! \overline{k}_{l}-i m g}
    {\Gamma_{j} \! - \! \overline{k}_{l}+i m g} \! = \!
    e^{i \overline{k}_{l} L} e^{-m \alpha L}, \\
    &\prod_{l=1}^{N} \!
    \frac{\Gamma_{j} \! - \! \overline{k}_{l}-i m g}
    {\Gamma_{j} \! - \! \overline{k}_{l}+i m g} \! = \! - 
    e^{-2 m \alpha L} \prod_{i'=1}^{N_\uparrow} \!
    \frac{\Gamma_{j} \! - \! \Gamma_{j'}-2 i m g}
    {\Gamma_{j} \! - \! \Gamma_{j'}+2 i m g},
    \label{BA equation 2}
    \end{aligned}
\end{equation}
where $\Lambda_{i}$ and $\Gamma_{j}$ are the quasi-momenta of spin-down and up particles respectively.
Eq. (\ref{BA equation}) and Eq. (\ref{BA equation 2}) yield the same solution to $\{ \overline{k}_l \}$.

In what follows, BA equations will be solved in the case of $L \gg \lambda_s$. 
Let us examine the first set of equations in Eq. (\ref{BA equation}).
Since $\Lambda_i-\overline{k}_l-i m g$ and $\Lambda_i-\overline{k}_l+i m g$ have the same magnitude, their quotient would not give an exponentially diverge factor $e^{L/\lambda_s}$, with two exceptions: 
(a) The denominator is nearly zero, {\it i.e.}, $\Lambda_i-\overline{k}_{i}+i m g \sim e^{-L/\lambda_s}$. 
(b) $\text{exp}(i \overline{k}_l L)$ cancels with the divergence factor $e^{L/\lambda_s}$, implying that the leading order of $\Im \overline{k}_l$ is $ m\alpha$. 
Based on this observation, we propose the following trial solution
\begin{equation*}
\begin{cases}
    \overline{k}_{i} = \Lambda_i + img - \epsilon_i e^{-2m\alpha L}, 
    &1 \leq i \leqslant N_\downarrow ; \\
    \overline{k}_{j} = \chi_{j}/L + i(m\alpha - \eta_{j}/L), 
    &N_\downarrow + 1 \leq j \leqslant N,
\end{cases} 
\end{equation*}
where $\epsilon_i$, $\chi_j$ and $\eta_j$ are dimensionless real numbers.
The same reasoning can be applied to Eq. (\ref{BA equation 2}), which yields another trial solution
\begin{equation*}
\begin{cases}
    \overline{k}_{i} = \chi_{i}/L - i(m\alpha - \eta_{i}/L),
    & 1 \leq i \leqslant N_\downarrow; \\
    \overline{k}_{j} = \Gamma_j - img - \epsilon_j e^{-2m\alpha L},
    & N_\downarrow + 1 \leq j \leqslant N.
\end{cases}
\end{equation*}
The two trial solutions should be consistent with each other. 
We combine the them into 
\begin{equation}
\small
\begin{cases}
    \overline{k}_{i} = \chi_{i}/L - i (m\alpha - \eta_{i}/L), \ \
    &\Lambda_i = \overline{k}_{i} - img + \epsilon_i e^{-2m\alpha L}; \\
    \overline{k}_{j} = \chi_{j}/L + i (m\alpha - \eta_{j}/L), \ \
    &\Gamma_j = \overline{k}_{j} + img + \epsilon_j e^{-2m\alpha L}.
\end{cases} \label{Assumption to k}
\end{equation}

By taking the trial solution Eq. (\ref{Assumption to k}) to BA equations Eq. (\ref{BA equation}) and Eq. (\ref{BA equation 2}) respectively, we obtain
\begin{equation}
\begin{cases}
    (1+g/\alpha)^{N_\uparrow} = e^{\eta_{i}} e^{i \chi_{i}}; \\
    (1+g/\alpha)^{N_\downarrow} = e^{\eta_{j}} e^{i \chi_{j}},
\end{cases}
\end{equation}
and
\begin{equation}
    \begin{aligned}
    -\frac{2img}{\epsilon_i} \prod_{i' \neq i}^{N_\downarrow}
    \frac{-2img}{\overline{k}_{i'}-\overline{k}_{i}} &=
    e^{-\eta_{i}} e^{i \chi_{i}}, \\
    \frac{\epsilon_j}{2img} \prod_{j' \neq j}^{N_\uparrow}
    \frac{\overline{k}_{j'}-\overline{k}_{j}}{2img} &=
    e^{\eta_{j}} e^{i \chi_{j}},
    \label{Further simplified first set BA equation}
    \end{aligned}
\end{equation}
which gives the following solution to $\{ \overline{k}_l \}$:
\begin{equation}
\! \! \! \! \! \! \!
\begin{cases}
    \overline{k}_{i} = 2 n_i \pi / L - i(m\alpha - N_\uparrow \eta_r / L); \\
    \overline{k}_{j} = 2 n_j \pi / L + i(m\alpha - N_\downarrow \eta_r / L).
    \label{BA solution to k_bar}
\end{cases} 
\end{equation}
$\eta_r$ defined in Eq. (\ref{Definition of kr}) describes the localization strength of the two-body case. 
Note that the sub-leading order of $\text{Im} \overline{k}_l$ will scale with the density. 
To prevent it from exceeding the leading order term $m\alpha$, the following constraint should be proposed
\begin{equation}
    \frac{N}{L} \ll \frac{m \alpha}{\ln (1+g/\alpha)},
\end{equation}
which is the dilute limit claimed in the paper.
We check our trial solution by taking it to the second set of equations in Eq. (\ref{BA equation}), which yields
\begin{equation*}
\begin{aligned}
    -\frac{2img}{\epsilon_i} (1+g/\alpha)^{N_\uparrow}
    \prod_{i' \neq i}^{N_\downarrow} 
    \frac{2img}{\overline{k}_{i'}-\overline{k}_{i}}=
    (-1)^{N_\downarrow-1}.
\end{aligned}
\end{equation*}
This equation matches the first equation in Eq.~(\ref{Further simplified first set BA equation}), confirming the correctness of the trial solution. 
The second equation in Eq.~(\ref{Further simplified first set BA equation}) can be recovered in parallel.

To determine the many-body wavefunction, we transform $\overline{\varphi}_\sigma\left(x_l\right)$ back to $\varphi_\sigma\left(x_l\right)$, where:
\begin{equation}
\begin{aligned}
    \varphi_\sigma\left(x_l\right)= 
    &\sum_{Q, P} \theta\left(x_{Q_1} > x_{Q_2} > \cdots > x_{Q_N} \right) \\
    &\times A_\sigma(Q, P) \ \exp \left(i \sum_l {k}_{P_l, \sigma_l} x_{Q_l}\right).
    \label{Bare BA wavefunction 2}
\end{aligned}    
\end{equation}
In this expression, $k_{P_l,\sigma_l} = \overline{k}_{P_l} - im\alpha\sigma_l$ with $\sigma_l = \pm 1$ representing the spin $z$-component. 
Without loss of generality, we assume $\sigma_i=-1$, $\sigma_j=1$ for any $1 \leq i \leq N_\downarrow$, $N_\downarrow + 1 \leq j \leq N$. 
For a given permutation $\{ P_l \}$ and $\{ Q_l \}$, $\{ k_{P_l,\sigma_l} \}$ defines a scattering channel, where $1 \leq P_l \leq N$. These scattering channels can be divided into two classes:
\begin{equation*}
\begin{aligned}
    \text{Class (a):} \quad \exists \ l, \
    \Im \ k_{P_{l}, \sigma_{l}} &= \pm (2m\alpha - N_{\downarrow, \uparrow}\eta_r/L), \\
    \text{Class (b):} \quad \forall \ l, \
    \Im \ k_{P_{l}, \sigma_{l}} &= \pm N_{\uparrow, \downarrow}\eta_r/L.
\end{aligned}
\end{equation*}
We will now prove that class (a) can be discarded in the dilute limit.
For a channel in class (a), if the momentum of a spin-down particle satisfies $\Im \ k_{P_{i}, \sigma_{i}} = 2m\alpha - N_{\uparrow}\eta_r/L$, there must exist a corresponding spin-up particle with momentum $\Im \ k_{P_{j}, \sigma_{j}} = -(2m\alpha - N_{\downarrow}\eta_r/L)$.
Class (a) can have many such up-down pairs, each pair contributing an exponential factor $e^{2(x_\uparrow-x_\downarrow)/\lambda_s}$. 
Depending on the coordinate permutation ($-L<x_\uparrow-x_\downarrow<0$ or $0<x_\uparrow-x_\downarrow<L$), this factor is either divergent or suppressed in the case of $L \gg \lambda_s$. 
If it diverges, the PBC can not be satisfied unless the corresponding scattering amplitude scales as $e^{-2L/\lambda_s}$. 
Therefore, regardless of whether the plane waves diverge or not, class (a) will always be suppressed by a factor $e^{-2x/\lambda_s}$ with $0<x<L$.
An exception occurs when $x < \lambda_s$, which means a pair of spin-up and spin-down particles get very close to each other. 
However, in the dilute limit, the average distance between particles $L/N$ is much larger than $\lambda_s$, rendering the exceptional region negligible. 
Furthermore, scattering channels of class (a) will not accumulate to a finite value as $N\rightarrow \infty$. 
Although the number of these channels increase as $N!$, they are not coherent and will cancel out each other. 
To justify this argument, we numerically calculate the normalization of the approximated wavefunction and the exact wavefunction in the 4-particle case with 2 up-down pairs.
They differ by a ratio of the magnitude $N\lambda_s/L$, which is roughly the normalization of a single channel in class (a).

In Fig.~\ref{4-particle wavefunction}, we plot the probability distribution function $\left| \varphi_{\downarrow \downarrow \uparrow \uparrow}\left(x_1, x_2, x_3, x_4 \right) \right|^2$ with fixed $x_2, x_3, x_4$. 
The red line represents the exact wavefunction, while the blue line represents the approximated wavefunction. 
From figures (a) to (f), a pair of spin-up and spin-down particles get progressively closer to each other. 
In figures (a) to (d), the approximated probability distribution almost overlaps with the exact one. 
They only diverge in the vicinity of the spin-up particle at $x_4$, where the exact probability is smaller. 
This implies that the scattering channels of class (a) contribute a negative part to the probability, preventing an up-down pair getting too close. 
Such approximation breaks down when $x_2-x_3 \sim \lambda_s$, as shown in figures (e) and (f). 
Nevertheless, the contribution to the normalization from the break-down region is negligible. 

With the reasoning above, all channels of class (a) can be discarded. 
In class (b), the solution Eq.~(\ref{BA solution to k_bar}) is written as:
\begin{equation}
\begin{cases}
    k_{i} = (2 n_i \pi + i N_\uparrow \eta_r) / L; \\
    k_{j} = (2 n_j \pi - i N_\downarrow \eta_r) / L.
    \label{Solution in class b}
\end{cases} 
\end{equation}
As we sum over all momentum permutations in the wavefunction Eq.~(\ref{Bare BA wavefunction 2}), only those channels of class (b) should be included. 
For any fixed coordinate permutation, different channels in class (b) can transform to each other by reflection between particles with identical spins, and the corresponding phase shift is $-1$ due to Pauli's exclusion principle. 
Consequently, these plane waves with momentum distribution Eq. (\ref{Solution in class b}) organize into a product of two Slater determinants for two spin components, simplifying the wavefunction to: 
\begin{equation}
\begin{aligned}
    &\varphi_\sigma\left(x_l\right)= 
    \text{det} (e^{i \chi_{i_1} x_{i_2} / L }) 
    \text{det} (e^{i \chi_{j_1} x_{j_2} / L }) e^{-\sum_{ij} \frac{1}{L} \eta_r(x_i-x_j) } \\
    &\times \sum_{Q} \theta\left(x_{Q_1} > x_{Q_2} > \cdots > x_{Q_N} \right) 
    A_\sigma(Q, Q) 
    \label{Bare BA wavefunction 3}
\end{aligned}    
\end{equation}
$A_\sigma(Q, Q)$ denotes the scattering amplitude when the momentum permutation $P$ is identical to the coordinate permutation $Q$. 
To progress, we need to figure out the phase shift between these scattering amplitudes. 
Generally, $A_\sigma(Q, P)$ satisfies:
\begin{equation}
\begin{aligned}
A_\sigma \left(Q, P \right) 
= \
&\frac{\overline{k}_{P_{l+1}}-\overline{k}_{P_l}}
{\overline{k}_{P_{l+1}}-\overline{k}_{P_l}+2i mg} A_\sigma \left(Q', P' \right) - \\
&\frac{2img}
{\overline{k}_{P_{i+1}}-\overline{k}_{P_i}+2i mg} A_\sigma \left(Q, P' \right),
\label{Scattering phase shift}
\end{aligned}
\end{equation}
where
\begin{equation*}
\begin{aligned}
    A_\sigma\left(Q', P' \right) &= 
    A_\sigma\left(\cdots Q_{l+1} Q_{l} \cdots, \cdots P_{l+1} P_{l} \cdots \right), 
    \\
    A_\sigma\left(Q, P' \right) &= 
    A_\sigma\left(\cdots Q_l Q_{l+1} \cdots, \cdots P_{l+1} P_l \cdots \right).
\end{aligned}
\end{equation*}
Here $A_\sigma(Q, P)$, $A_\sigma(Q', P')$, $A_\sigma(Q, P')$ represents the amplitudes of incident, transmission and reflection waves respectively. 
There are two cases to consider:
\begin{align*}
&\text{(A): } 1 \leq Q_l, \ Q_{l+1} \leq N_\downarrow \text{ or } N_\downarrow + 1 \leq Q_l, \ Q_{l+1} \leq N \\
&\text{(B): } 1 \leq Q_l \leq N_\downarrow \text{ and } N_\downarrow + 1 \leq \ Q_{l+1} \leq N
\end{align*}
In case (A) $A_\sigma(Q, Q)=A_\sigma(Q', Q')$ due to the Fermi statistics. 
Case (B) represents the scattering between an up-down pair. According to Eq.~(\ref{Scattering phase shift})
\begin{equation*}
A_\sigma \left(Q, Q \right) = 
\frac{1}{1 + g/\alpha} A_\sigma \left(Q', Q' \right) - 
\frac{g/\alpha}{1 + g/\alpha} A_\sigma \left(Q, Q' \right),
\end{equation*}
where all the phase shifts are kept to the leading order.
The reflection channel belongs to class (a), with the corresponding amplitude $A_\sigma(Q, Q')\sim e^{-2L/\lambda_s}$, which vanishes in the dilute limit. 
Thus, we arrive at
\begin{equation}
    \frac{A_\sigma(Q', Q')}{A_\sigma(Q, Q)} = 1 + \frac{g}{\alpha} = e^{\eta_r}.
    \label{up-down pair phase shift}
\end{equation}
Such phase shift applies to any up-down pair and is independent of the relative momentum between the pair. 
With this setup, the wavefunction Eq.~(\ref{Bare BA wavefunction 3}) can be further simplified.
First we define
\begin{equation*}
\begin{aligned}
    &\sum_{Q} \theta\left(x_{Q_1}<x_{Q_2}<\cdots<x_{Q_N} \right) A_\sigma(Q, Q) \\
    &= A_\sigma(x_1, x_2, \cdots x_N)
\end{aligned}
\end{equation*}
with 
\begin{equation*}
\begin{aligned}
    & A_\sigma(x_1, x_2, \cdots x_N) = 1 \\ 
    & \text{if} \ \ 
    x_{N_\downarrow+1}<x_{N_\downarrow+2}<\cdots<x_N
    <x_1<x_2<\cdots<x_{N_\downarrow}.
\end{aligned}
\end{equation*}
The function $A_\sigma(x_1, x_2, \cdots x_N)$ is a constant for a given coordinate permutation.
Its value changes only when the coordinates of an up-down pair switch.
When all the spin-up particles are on the left side of spin-down particles, its value is defined as 1.
According to the constant phase shift Eq. (\ref{up-down pair phase shift}), every time a spin-up particle cross a spin-down particle from the left to the right, the value of $A_\sigma(x_1, x_2, \cdots x_N)$ will multiply a factor $e^{-\eta_r}$.
Therefore, we can use the following algorithm to determine the output of this function:

{\it Step 1: Find all the positions of the spin-up particles}

{\it Step 2: For every spin-up particles, count how many spin-down particles staying on its left side. This number is denoted as $u_j$, where $j$ is the index of spin-up particles.}

{\it Step 3: Sum over $u_j$. The value of the function is given by $e^{-\eta_r \sum_j u_j}$. }

$u_j$ can also be expressed as
\begin{equation}
    u_j = \sum_i^{N_\downarrow} \theta(x_j - x_i).
\end{equation}
In other words,
\begin{equation}
    A_\sigma(x_1, x_2, \cdots x_N) = e^{-\eta_r \sum_{ij} \theta(x_j-x_i)},
\end{equation}
which simplifies Eq. (\ref{Bare BA wavefunction 3}) to
\begin{equation*}
    \varphi_\sigma\left(x_l\right)  =  
    \text{det} (e^{i \chi_{i_1} x_{i_2}/L }) 
    \text{det} (e^{i \chi_{j_1} x_{j_2}/L }) 
    e^{-\frac{1}{2} \sum_{i j} \!  W(x_{i}-x_{j})}.
\end{equation*}
This is the wavefunction shown in the main body of the paper.


\begin{figure}[tp]
    \centering
    \includegraphics[scale=0.36]{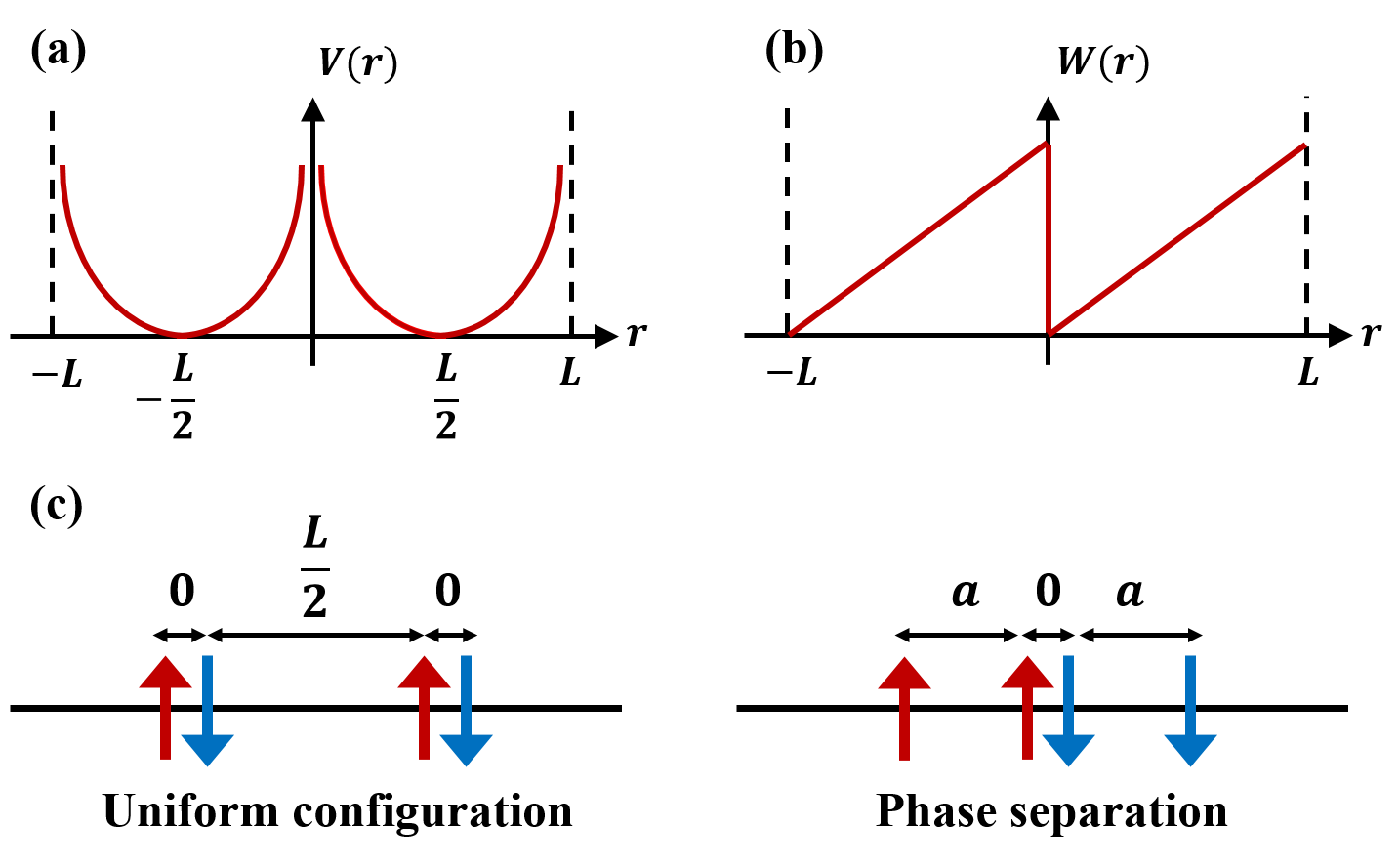}
    \caption{($a$) and ($b$) shows the Pauli repulsion energy $V(r)$ and the resonance energy $W(r)$ respectively. 
    They are periodical functions due to the periodical boundary condition.
    ($c$) illustrates the configurations with the locally minimal energy in the zero temperature limit.
    In the first one, four particles form two up-down pairs with distance $L/2$. 
    The size of the pair is negligible; 
    In the second one, particles with identical spins form two adjacent clusters with size $a= L  \tan^{-1} (\frac{\pi}{2} \eta_r)/\pi$.
    }
    \label{fig: 4-particle Phase Transition}
\end{figure}

\section{III. Phase transition in the four-particle case} \label{Phase transition in the four-particle case}

In this section, we calculate the transition point for the system with two up-down pairs.
The probability distribution function $\rho_\beta$ is defined as
\begin{equation}
    \rho_\beta = e^{-\beta \mathcal{H}},
\end{equation}
where $\mathcal{H}$ is a sum over the Pauli repulsion energy $V(r)$ between particles of identical spins and the resonance energy $W(r)$ between particles of different spins.
The shape of $V(r)$ and $W(r)$ is illustrated in Fig. \ref{fig: 4-particle Phase Transition} ($a$) and ($b$) respectively.
In the zero-temperature limit $\beta \rightarrow \infty$, the system freezes into the state with minimal energy. 
For the four-particle case, $\mathcal{H}$ has two local minima, whose configurations are shown in Fig. \ref{fig: 4-particle Phase Transition}. 
The energies of the uniform configuration $E_1$ and the phase separation $E_2$ are given by:
\begin{eqnarray}
\left 
\{
\begin{array}{l}
E_1 = 2 \eta_r, \vspace{1ex} \\
E_2 = \frac{8\eta_r a}{L} - 4 \ln ( \sin \frac{\pi a}{L}). \vspace{1ex}
\end{array}
\right.
\end{eqnarray}
As a function of the cluster size $a$, $E_2$ is minimized when $a= L \tan^{-1} (\frac{\pi}{2} \eta_r)/\pi$. 
The transition point is determined by setting $E_1=E_2$, which gives $\eta_r  = 3.84$.
One finds the energy minimum switch from $E_1$ to $E_2$ as crossing the transition point, which means the uniform configuration turns to phase sepration, as prediction.


\begin{figure}[tp]
    \flushleft
    \includegraphics[scale=0.35]{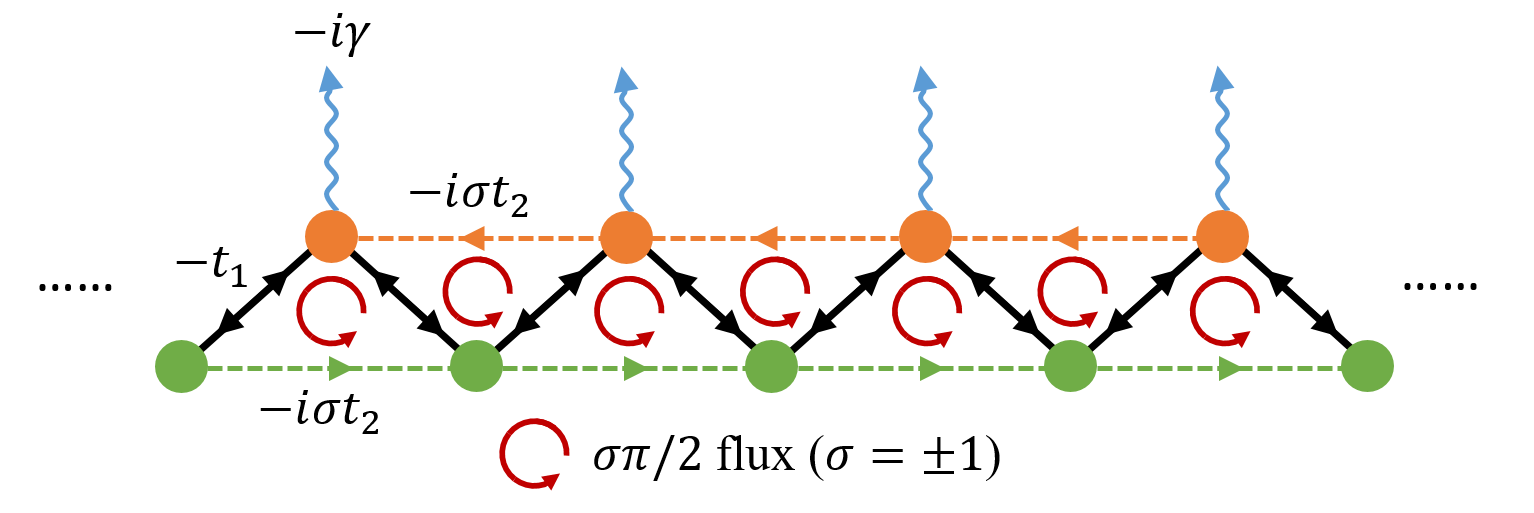}
    \caption{Lattice chain corresponding to the Hamiltonian $\hat{H}_L$.
    Each unit cell of the chain consists of two sublattice sites, denoted by green and orange dots respectively. 
    Dissipation is introduced on the orange sites via coupling to a reservoir, which is modeled by the on-site loss term $-i\gamma c_n^\dagger c_n$, while the green sites are free of loss. 
    A spin-dependent flux $\phi=\sigma \pi/2$ threads each triangular plaquette formed by the green and orange sites, contributing to a phase shift to the next nearest neighbour hopping.
    $\sigma=\pm 1$ represents the spin $z$-component. }
    \label{LatticeModel}
\end{figure}

\section{IV. Experimental realization}
\label{Experimental realization}

\begin{figure*}[tp]
    \flushleft
    \includegraphics[scale=0.365]{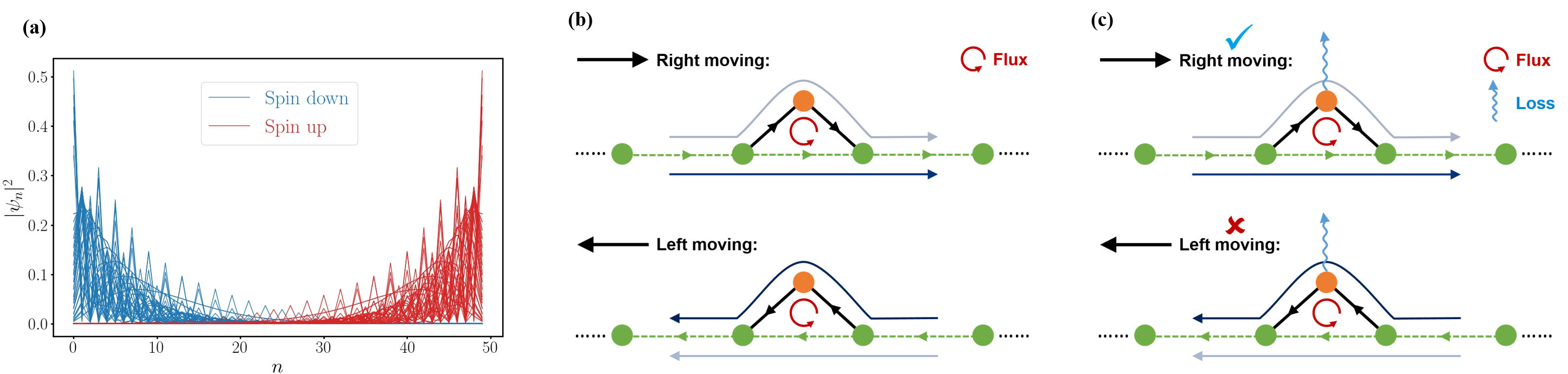}
    \caption{($a$) shows all the single-body eigenstates of the lattice Hamiltonian $\hat{H}_L$ upon the OBC.
    $n$ and $\psi_n$ represent the index of the lattice site and the value of the eigenstate $\psi$ at this site, respectively. 
    As illustrated, spin-up and down particles localize at different boundaries.
    Here $t_1=1$, $t_2=0.1$ and $\gamma=1$.
    ($b$) and ($c$) shows the moving patterns of spin-up particles ($\sigma=1$) on the lattice defined by $\hat{H}_l$. 
    ($b$) Right and left moving modes tend to take different paths due to the magnetic flux. 
    ($c$) The loss on the orange site does not affect the main path of the right moving mode, while blocks that of the left moving mode.}
    \label{fig: Lattice model}
\end{figure*}

As for experimental realizations, consider the following 1D lattice Hamiltonian
\begin{eqnarray}
\begin{aligned}
\hat{H}_L & = - \sum_{n,\sigma}
\left( 
t_1 c_{n, \sigma}^{\dagger} c_{n+1, \sigma} 
- it_2 (-1)^n  c_{n, \sigma}^{\dagger} \sigma^z c_{n+2, \sigma} 
+ \text{h.c.} 
\right) \nn \\
& - i\gamma\sum_{n \text{ odd}, \ \sigma}(-1)^n c_{n, \sigma}^{\dagger}c_{n, \sigma}. 
\label{Lattice Hamiltonian}
\end{aligned}
\end{eqnarray}
Here $n$ and $\alpha$ of the creartion operator $c_{n, \alpha}^{\dagger}$ are the indices for site and spin, respectively.
$t_1$, $t_2$, $\gamma$ are all real, with $|t_1|>|t_2|$. 
$t_2$-term stands for the Hermitian SOC; 
$\gamma$-term represents the loss on the odd index sites, which can be realized in the cold atomic systems \cite{Yanbo2022ColdAtom}.
Notice that $H_L$ is free of gain, in which case the dynamics described the Lindbladian can be fully determined by its corresponding non-Hermitian Hamiltonian \cite{buvca2020bethe}. 
In Fig. \ref{LatticeModel} we illustrate the corresponding lattice chain with green and orange dots denoting different sublattice sites.

The lattice Hamiltonian $\hat{H}_L$ and the continuum one $\hat{H}$ are equivalent in the sense that $\hat{H}_L$ exhibits an effective imaginary spin-orbit coupling (SOC). 
One can check this in the momentum representation of $\hat{H}_L$, given by
\begin{equation}
\mathcal{H}_L(k)=\left(\begin{array}{cc}
2 t_2 \sigma^z \sin k & t_1\left(1+e^{-i k}\right) \\
t_1\left(1+e^{i k}\right) & -2 t_2 \sigma^z \sin k-i \gamma
\end{array}\right),
\end{equation}
Here $\mathcal{H}_L(k)$ is written in the sublattice space.
The eigen-value of $\mathcal{H}_L(k)$ is
\begin{equation*}
E_k = - i\frac{\gamma}{2} \pm \sqrt{ \left(2t_1 \cos \frac{k}{2} \right)^2 + \left(2t_2 \sin k + i \frac{\gamma}{2}\sigma \right)^2},
\end{equation*}
where $\sigma=\pm 1$ represents the spin $z$-component.
In the long-wavelength limit, $E_k$ can be expanded in powers of $k$ up to second order:
\begin{equation}
E_k=-i \frac{\gamma}{2} \pm\left(\epsilon_0+\frac{k^2+2 i m \alpha \sigma k}{2 m}\right),
\end{equation}
with
\begin{equation*}
m=\frac{\epsilon_0}{t_1^2}\left[\left(\frac{4 t_2}{\epsilon_0}\right)^2-1\right]^{-1}, \quad \alpha=\frac{\gamma t_2}{\epsilon_0}, \quad \epsilon_0=\sqrt{4 t_1^2-\frac{\gamma^2}{4}} .
\end{equation*}
In this limit, the lattice Hamiltonian $\hat{H}_L$ reduces to the continuous Hamiltonian $\hat{H}$, where the term $2i m\alpha \sigma k$ in $\hat{H}_L$ captures the non-Hermitian SOC. 
Although this non-Hermitian term contributes an unbounded imaginary part to $E_k$ at the level of perturbative expansion, the exact non-perturbative expression of $E_k$ indicates that $\text{Im}(E_k)<0$ for all $k$, implying that the system intrinsically has no gain.
Furthermore, exact diagonalization of $\hat{H}_L$ in the case of open boundary condition shows the spin-dependent NHSE, i.e., eigenstates of spin-up and down particles localize at different boundaries. 
This is shown in Fig \ref{fig: Lattice model} ($a$). 

The spin-dependent NHSE in this lattice model can be explained as following.
We start from a tight bonding chain with a single triangular cell in the middle, and take $\sigma=1$. 
As illustrated in Fig \ref{fig: Lattice model}. ($b$), when the orange site is free of loss, the Hamiltonian of the system is
\begin{equation*}
\begin{aligned}
&\hat{H_l} = \hat{H}_0 + \hat{H}_c \ , \quad
\hat{H}_0 = - \sum_{n} it_2 (c_{n}^{\dagger} c_{n+2}  - c_{n+2}^{\dagger} c_{n}) \ , \\
&\hat{H}_c = -t_1(d^\dagger c_m + d^\dagger c_{m+1} + \text{h.c.}) \ ,
\end{aligned}
\end{equation*}
where $\hat{H}_0$ represents the Hamiltonian of the tight bonding chain composed by the green sites, and $\hat{H}_c$ describes the coupling between the chain and the orange site, with $d^\dagger$ denoting the creation operator on that site. 
Under this setup, due to the interference of the wavefunction, right moving plane waves will have lower occupation on the orange site than that of the left moving ones. 
For example, a right moving plane wave with momentum $k=\pi$ does not pass through the orange site. 
It moves to the right because $v_k=dE_0/dk$ is positive at $k=\pi$, where $E_0$ is the eigen-value of $\hat{H}_0$; 
and its occupation vanishes on the orange site because the hoppings from $m$th and $(m+1)$th sites cancel with each other. 
On the contrary, coherent interference happens on the orange site for left moving plane waves, for instance with $k=0$. 
Therefore, when the orange site is subjected to loss, right moving waves can avoid it while left moving waves can not, such that the life time of the former will be longer than that of the latter. 
Consequently, spin-up particles will accumulate on the right boundary upon the OBC. 
This is illustrated in Fig. \ref{fig: Lattice model} ($c$). 
The reverse skin effect happens for spin-down particles, which can be understood by performing a parity transformation to $\hat{H}_l$.
The lattice model given by $\hat{H}_L$ is obtained by repeating the triangular cell described above. 
In this case, all the long-live modes mainly distribute on the green chain to avoid the loss on the orange chain, where spin-up and down particles preferring right and left hopping respectively. 
This induces an effective non-Hermitian spin-dependent hopping, giving rise to the spin-dependent NHSE. 

With the imaginary SOC, it can be expected that $\hat{H}_L$ will also exhibits many-body resonance effect if the repulsive interaction is turned on. 
Due to the appearance of the next-nearest-neighbor hopping in $\hat{H}_L$, such a system is no longer integrable.  
Nevertheless, our results via BA provide a good starting point for further exploring the exotic physics on the interplay between strong interaction and non-Hermitian physics.

The resonance effect can be experimental detected by studying the dynamics of particles, which is similar to the observation of single-body NHSE. 
In the single-body case, a spin-up particle initially placed in the middle of the lattice chain will propagate to the right side due to the NHSE.
If a spin-down particle is added on its right side, the pair will form a stable state in the bulk instead of propagating to the opposite sides, which can be seen as an evidence for the resonance effect.
By placing many spin-up and down particles on the chain, the two groups of particles will mix when the repulsive interaction is weak and become spatially separated if the interaction strength exceeds a critical value,  which is the signal for the phase transition.

\end{document}